\documentclass{elsart}
\usepackage{epsfig}
\usepackage{amssymb}

\begin{document}
\begin{frontmatter}

\title{Efficient Promotion Strategies in Hierarchical Organizations } 

\author
{Alessandro Pluchino} \ead{alessandro.pluchino@ct.infn.it}
\address{Dipartimento di Fisica e Astronomia,  Universit\'a di Catania,\\
and INFN sezione di Catania,  Via S. Sofia 64, I-95123 Catania, Italy} 

\author
{Andrea Rapisarda} \ead{andrea.rapisarda@ct.infn.it}
\address{Dipartimento di Fisica e Astronomia,  Universit\'a di Catania,\\
and INFN sezione di Catania,  Via S. Sofia 64, I-95123 Catania, Italy} 

\author
{Cesare Garofalo}  \ead{cesaregarofalo@yahoo.com}
\address {Dipartimento di Sociologia e Metodi delle Scienze Sociali, Universit\'a di Catania, \\
Via Vittorio Emanuele II 8, I-95131 Catania, Italy}

%
%

\begin{abstract}
The  Peter principle has been recently investigated by means of an agent-based simulation and   its 
validity has been numerically corroborated. It has been confirmed that,  within certain conditions,  it  can  really influence in a negative way the efficiency of a pyramidal organization adopting meritocratic promotions.  It was  also found that, in order to bypass these effects, alternative promotion strategies should be adopted, as for example a random selection choice. In this paper, within the same line of research,  we  study promotion strategies in a more realistic  hierarchical and modular organization and we show the robustness of our previous results, extending their validity to  a more general context. We discuss also why the adoption of these strategies could be useful for real organizations. 

\end{abstract}

\end{frontmatter}

\section{Introduction}

Promotion strategies are fundamental for a hierarchical organization, being this a scientific group, a company, a public administration, a cluster of computers or a group of  animals. They are important  to understand the dynamics of a pyramidal system and  eventually provide ways  to improve its efficiency. It is not strange that also physicists are working in this direction. In fact in the last years physicists have started  to collaborate with economists and social scientists in order to get a more quantitative understanding of social sciences mechanisms \cite{Nature,Buchanan1,Buchanan2,Hedstrom1,Hedstrom2,Hedstrom3}. Actually, it is by now largely accepted that, even in social sciences, simple schematic models and computer simulations inspired by statistical physics are able to take into account unexpected collective behaviors of large groups of individuals, discovering emergent features independent of their individual psychological attributes, which are very often counterintuitive and difficult to predict just following common sense. Along these lines, by means of an agent-based simulation 
approach \cite{Epstein1,Epstein2,Miller,Castellano,Wilensky}, we study here the effects of the Peter 
principle \cite{Peter0} within a very general context where different promotions strategies are investigated in order to maximize the global efficiency in a given hierarchical system. In a paper published before 
\cite{Peter1} we have already studied this phenomenon within a pyramidal organization, showing its validity under certain conditions, and we have tested several strategies in order to bypass its negative effects. In this paper   we investigate in deeper detail a more complex modular organization, also endowed of new realistic features, in order to test these different promotion strategies under the Peter hypothesis and their influence in maximizing  the global efficiency of the system, considering also  the individual expectations of its members in terms of career progressions. In particular we study the gain in efficiency due to both the organization topology (modular or pyramidal) and the introduction of a variable percentage of random promotions after a meritocratic transient.   
The paper is organized as follows. In section $2$, after a brief summary of  our previous results, we present the details of the more realistic new model adopted in the present paper and we  compare the old pyramidal topology with the new modular one. In section $3$ we describe the new  simulation results. Then a general discussion is addressed in section $4$, where the real applicability of these strategies is also presented and finally some conclusions are drawn in section $5$.

\section{The extended hierarchical organization model}

We present in this section the details of the new hierarchical organization model adopted in this paper, in order to test in more realistic situations the effects of the Peter principle and the possible strategies to contrast it. However, before illustrating the new model, we summarize the results obtained in our  previous paper \cite{Peter1}.

\subsection{Previous results about the Pyramidal model}

In our previous paper we studied the  schematic pyramidal model shown in   Fig.1.   
The main features of this model are summarized below. 

\begin{figure}  
\begin{center}
\epsfig{figure=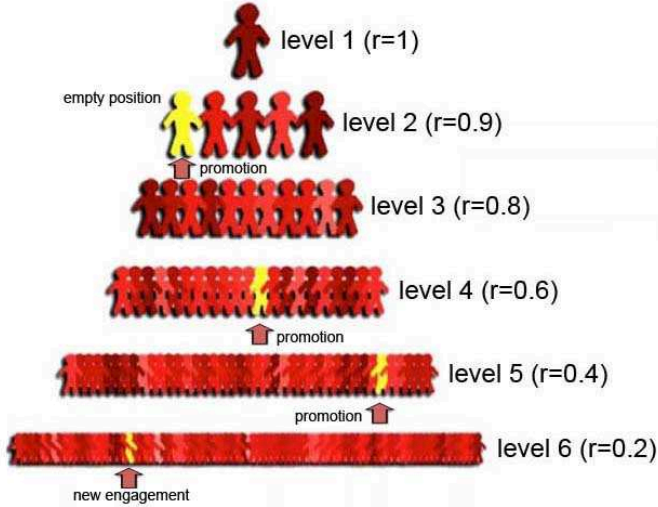,width=9truecm,angle=0}
\end{center}
\caption{ {\it Schematic view of the simple pyramidal model studied in our previous paper\cite{Peter1}. Beside each level, the corresponding value of responsibility (increasing linearly from the bottom to the top) is also reported. We 
show in yellow the empty positions for which promotions are required}.  
}
\label{organization}
\end{figure}

\begin{enumerate}
\item We considered a hierarchical pyramidal organization with 160 positions divided into six levels. Each level has a different number of members (which decreases, climbing the hierarchy) with a different responsibility, i.e. with a different weight on the global efficiency (see below) of the organization. The members of the organization  have only two features: an age (ranging from 18 to 60 years) and a degree of competence (ranging from 1 to 10 and indicated by a color with intensity proportional to the competence). As initial conditions we selected ages and competences following normal distributions with, respectively, average 25 (with standard deviation 5) and average 7 (with standard deviation 2);
\item At each time step, members with an age over the retirement threshold ( fixed at  60 years) or with a competence lower than the dismissal threshold (fixed at 4) leave the organization (their color becomes yellow) and someone from the level immediately below (or from outside for level 6) will be chosen for promotion, see Fig.1; 
\item	Four different competing strategies of promotions have been proposed. A first strategy consists in promoting the best worker, a second one in promoting the worst, a third one considers the promotion of  a random worker and a fourth one alternate the promotion of  the best and the worst;
\item	For each promotion at the upper level, two different mechanisms of competence transmission have been considered.
1) ÒCommon Sense (CS)Ó: if the features required from one level to the upper are enough stable, the new competence at the upper level is correlated with the previous one and the agent maintains his competence with a small error;
2) ÒPeter Hypothesis (PH)Ó: if the features required from one level to the upper can change considerably, the new competence at the upper level is NOT correlated with the previous one, so the new competence is again  randomly assigned  from a normal distribution, as happens in a new engagement; 
\item	A parameter, called global efficiency $E$, is calculated by summing the competences of the members level by level, multiplied by the level-dependent factor of responsibility, ranging from 0 to 1 and linearly increasing on climbing the hierarchy. The result is normalized to its maximum possible value $ Max (E)$ and to the total number of agents N, so that the global efficiency (E) can be expressed as a percentage. Therefore, if $C_i$ is the total competence of level $i$-th, the resulting expression for the global efficiency is 
\begin{equation}  
 E(\%)=\frac{\sum_{i=1}^6 C_i r_i}{Max(E) \cdot N} \cdot 100  ,
\label{eq0}
\end{equation} %
 being $Max(E)=\sum_{i=1}^6 (10 \cdot n_i) \cdot r_i / N$, where $n_i$ is the number of agents of level $i$-th
\end{enumerate}

\begin{figure}  
\begin{center}
\epsfig{figure=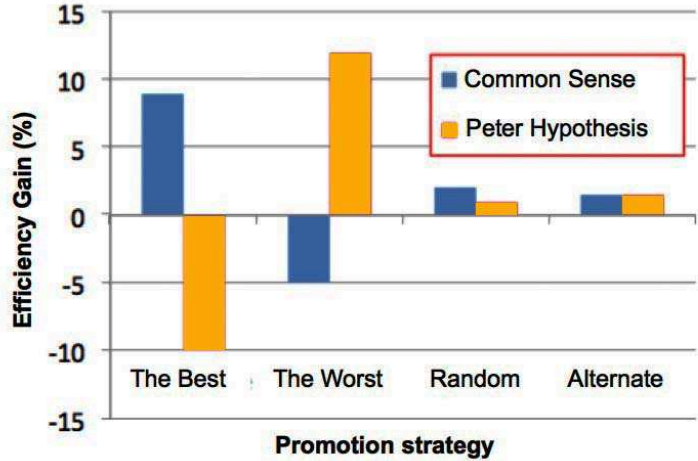,width=10truecm,angle=0}
\end{center}
\caption{ {\it A schematic summary of the results found in our previous paper \cite{Peter1}, see text}.  
}
\label{results1}
\end{figure}

The main results found in our previous paper are summarized in the histogram of Fig.2, where the asymptotic efficiency gain, calculated with respect to a common but arbitrary initial state and averaged over many different realizations of the initial conditions, is reported as function of the hypothesis adopted and of the promotion strategy applied.   
In particular we found that 
\begin{enumerate}
\item promoting the best member is a winning  strategy only if the CS hypothesis holds, otherwise is a loosing one - for this reason we also define the strategy "The best" coupled with the Peter hypothesis as "naively" meritocratic;
\item  on the contrary, if the PH holds, the best strategy is that one of promoting the worst member;
\item  but if one {\it does not know} which of the two hypothesis holds, then adopting a random promotion strategy, or alternating the promotion of the best and the worst, results to be always a winning choice. 
\end{enumerate}

The previous results are not straightforward nor immediately intuitive, since at a first sight they seem to contradict common sense; moreover they were obtained through a very simple toy model which could seem a very peculiar example. Therefore, although the paper was very successful and appreciated also for its simplicity  - it was quoted by several blogs and specialized newspapers, among which the MIT blog, the New York Times and the Financial Times, and it was also awarded the IG Nobel prize 2010 for "Management" \cite{ignobel} -  further investigations within a more realistic model were certainly desirable. 

\begin{figure}  
\begin{center}
\epsfig{figure=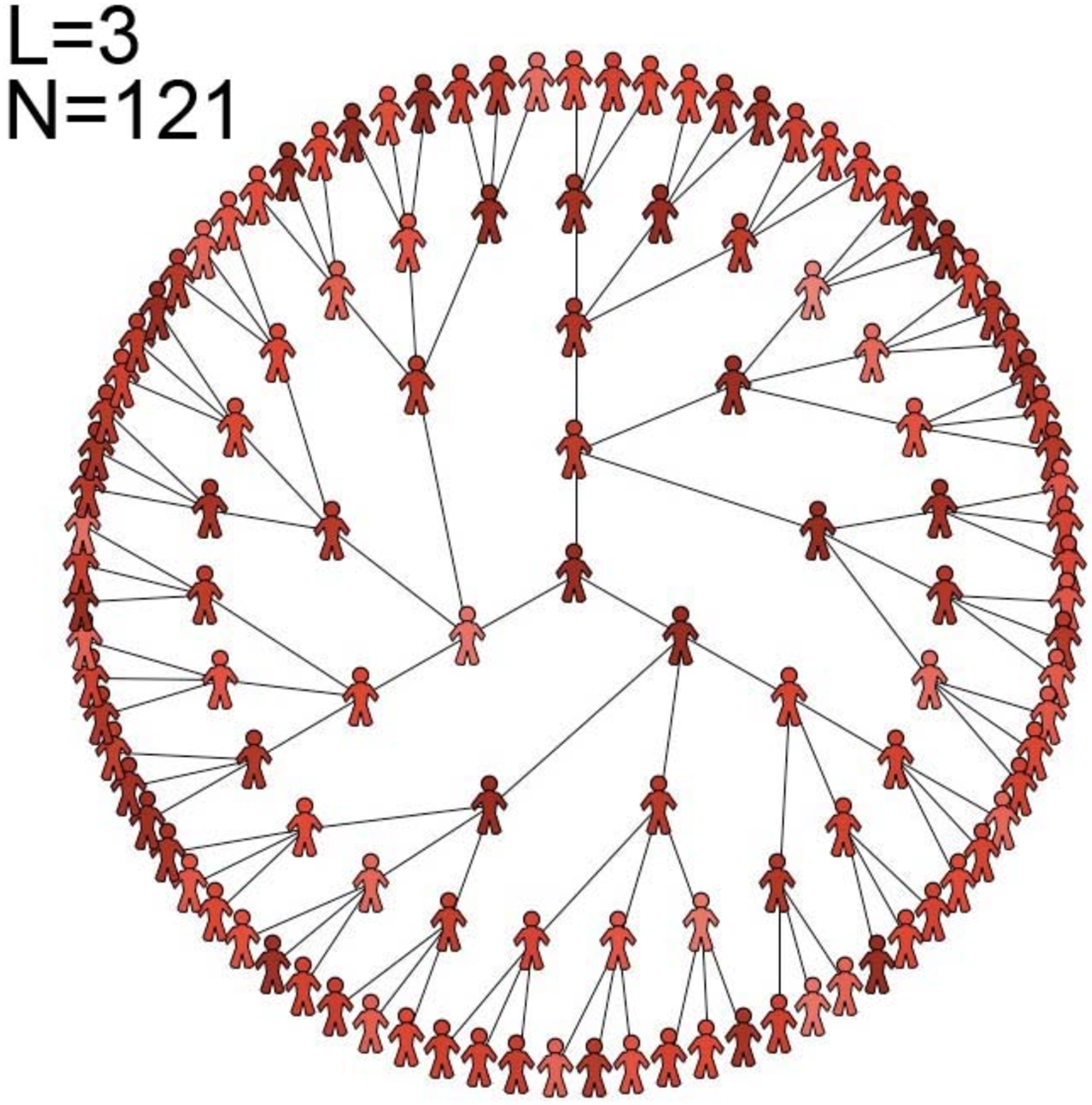,width=5truecm,angle=0}
\epsfig{figure=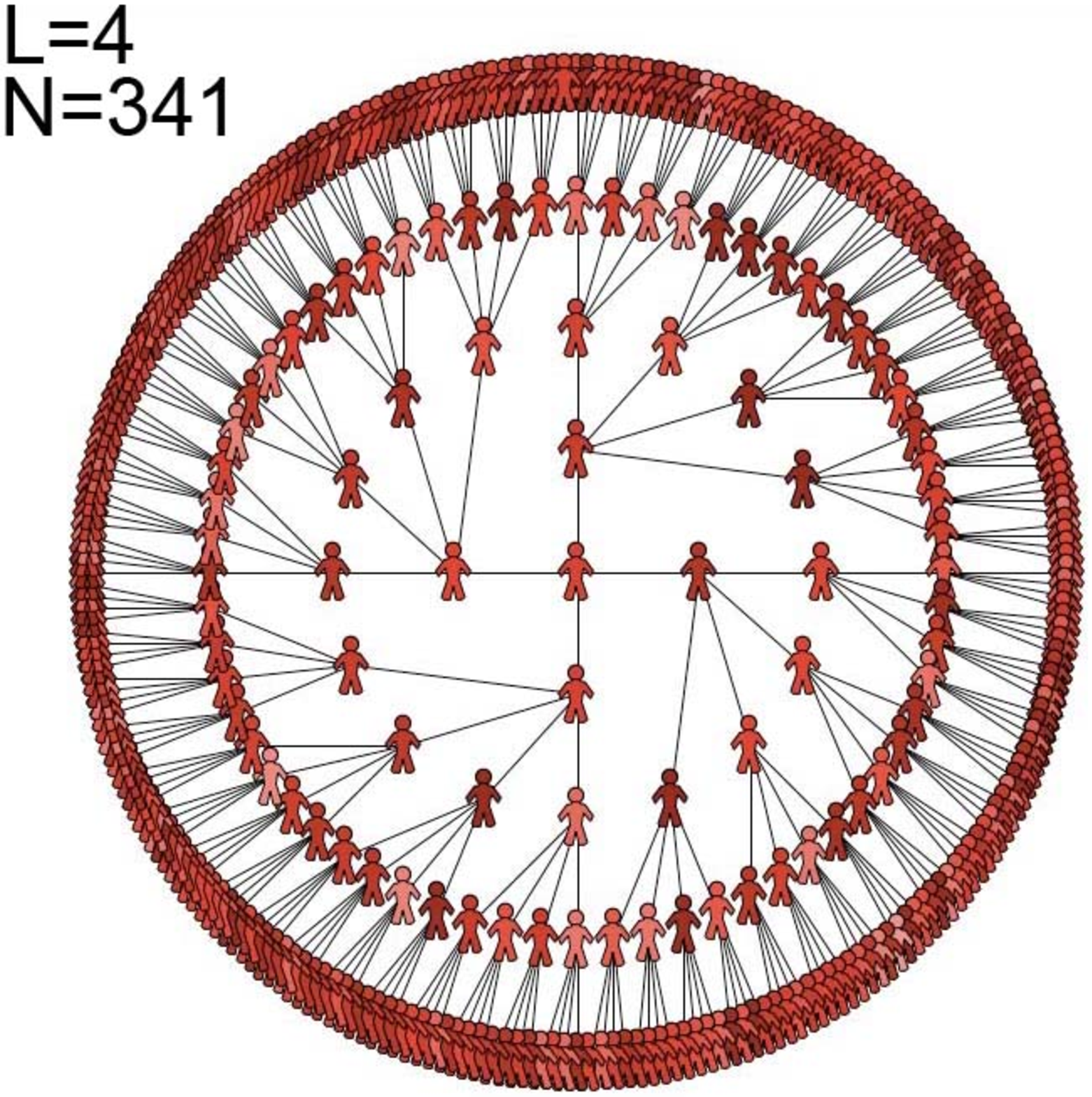,width=5truecm,angle=0}
\epsfig{figure=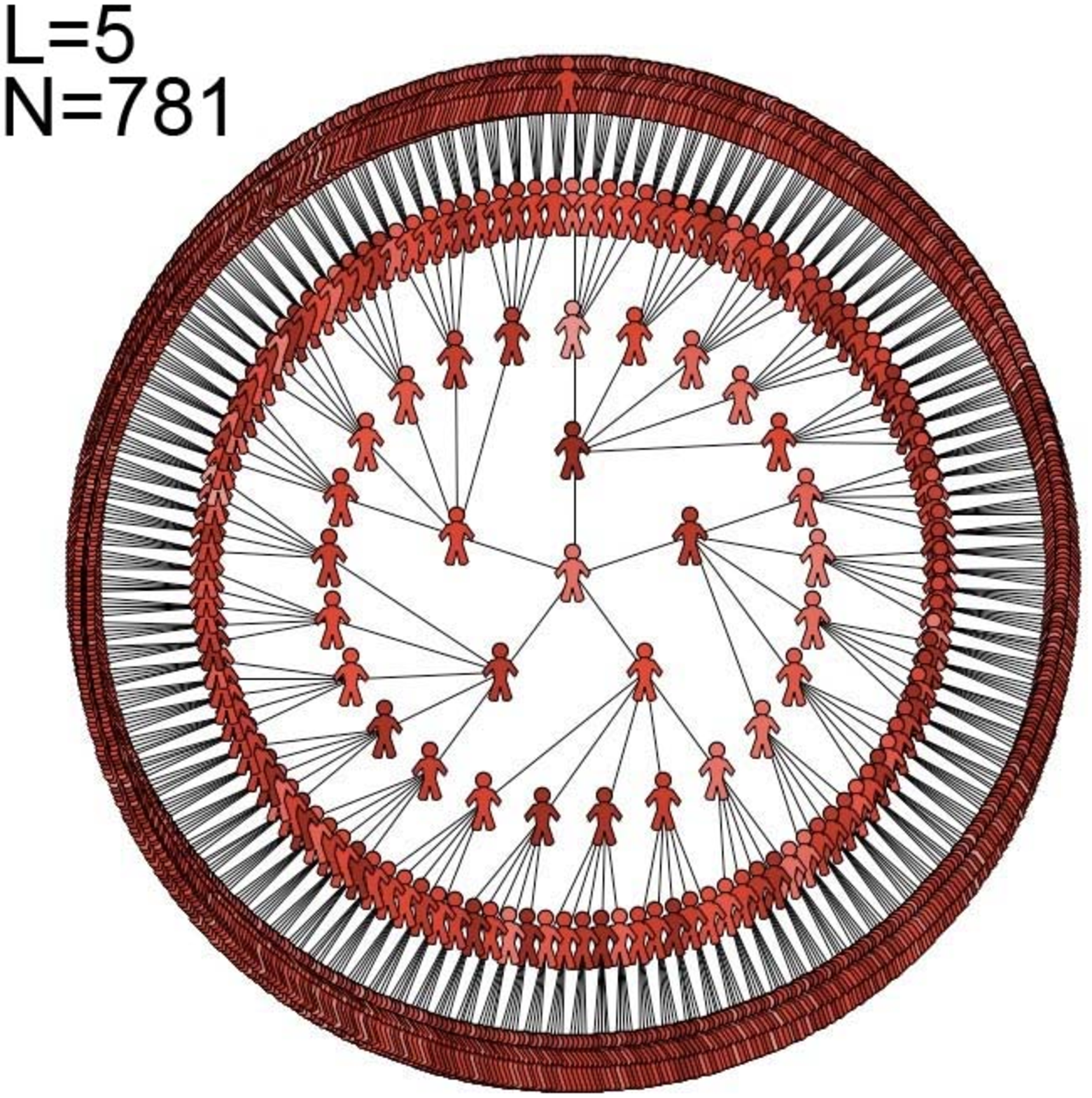,width=5truecm,angle=0}
\epsfig{figure=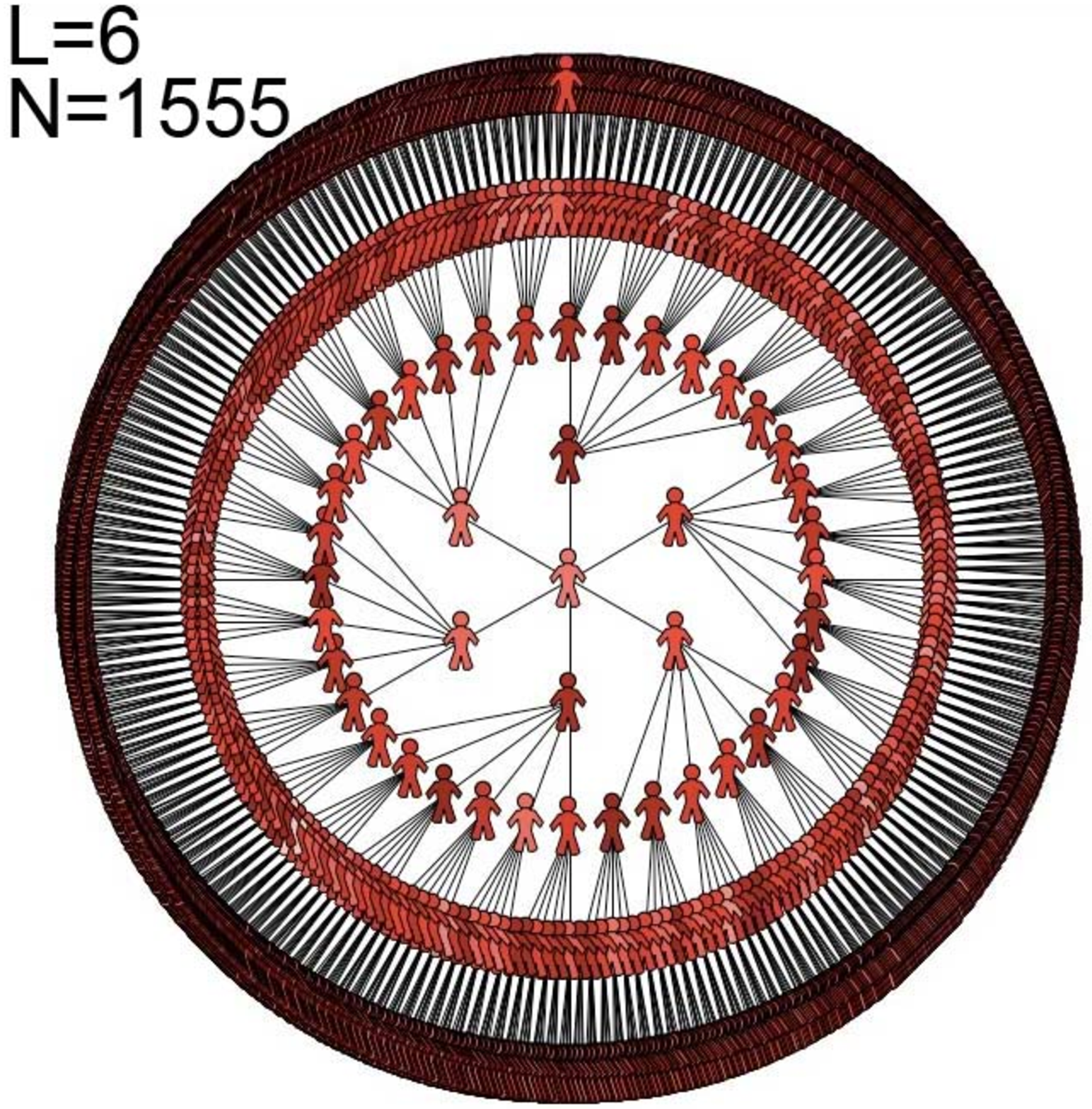,width=5truecm,angle=0}
\end{center}
\caption{ {\it Hierarchical tree networks with different coordination number $L$ and size $N$: for $L=3$ $N=121$, for $L=4$ $N=341$, for $L=5$ $N=781$ and for $L=6$ $N=1555$. } }  
\label{newmodel}
\end{figure}

\subsection{The new Hierarchical Tree model}

In order to test the robustness and the general validity of the increase in efficiency triggered by random strategies, we improved our agent based model by introducing several new features which provide a more realistic scenario. 
\\
First of all we consider a more complex topology for modeling a schematic modular organization, i.e. a hierarchical tree network with $K=5$ levels, where each agent (node) at levels $k=1,2,3,4$ (excluding the bottom level with $k=5$) has exactly $L$ first subordinates (i.e. first neighbors at level $k+1$), which will fill that position when it will become empty. An analogous structure has been used in a recent paper concerning a Dilbert-Peter model of organization effectiveness \cite{Peter-pawel}. We can call $L$ the {\it coordination number} of the network. This means that, at variance with the pyramidal schematic model of our first paper, in this case promoted agents follow the links to ascend through levels. On the other hand, neglecting the links and promoting agents from the entire level $k$ to the next level $k-1$, one recovers the pyramidal model as a particular case. For a given value of $K$ and $L$, the total number $N$ of agents of such a hierarchical network is given by 

\begin{equation}
N=\frac{L^K-1}{L-1}. 
\end{equation}

In the following we fix $K=5$ and vary $L$. In this case, each agent with $k<5$ has exactly $N_k=(L^{6-k}-L)/(L-1)$ subordinates in all the levels below.  
In Fig.3 we show four examples of hierarchical tree networks with $K=5$ levels and an increasing coordination number $L$: for $L=3$ we obtain a network with $N=121$ agents, for $L=4$ $N=341$, for $L=5$ $N=781$ and for $L=6$ $N=1555$. The responsibility value is $0.2$ for the bottom level and increases linearly, like in the previous model, with step $0.2$ for each level up to the top one, whose responsibility value is $1$.
Such a modular structure is surely more suitable than the simple pyramidal one to describe realistic sales divisions in large corporations or project teams in government institutions \cite{Peter-pawel}.      
\\
The second main improvement concerns the time units adopted in the simulations. In our previous model we adopted one year as time unit, therefore all the dynamical features of the algorithm (retirements, dismissals, promotions or new engagements) were updated at the end of each year. Now we enlarge the time resolution by using one month as unit. This means that the age of all the agents increases of one unit (one year) every $12$ time steps, but all the dynamical features are now updated every month. In such a way we will be able to follow the dynamics of the organization over a more realistic and detailed  time scale. 
\\
In order to make the dynamics independent of the initial conditions, instead of studying gains and losses with respect to an arbitrary initial state for the organization (with an arbitrary value of the initial global efficiency, as done in \cite{Peter1}), we will refer here to a {\it relative global efficiency} $E_r(\%)$, calculated with respect to a fixed transient during which a naively meritocratic strategy (i.e. promotions of "The best" coupled with Peter hypothesis) is always applied. The introduction of such a transient also allows us to simulate organizations which have already reached their stationary state, since we are not particularly interested in newly founded companies but in existing average-size companies.
\\ 
Another important improvement concerns the details of random promotions. Since it is understandable that a real company would initially be hesitant in adopting {\it tout court}  random promotions as a long term strategy, it is worthwhile to investigate how much randomness is effectively needed in order to see {\it as soon as possible}  some relevant improvement after the meritocratic transient. Therefore we introduce a new "Mixed" strategy, where a different increasing percentage of random promotions with respect to "The best" one is considered. Of course a "Mixed" strategy with $0\%$ of random promotions corresponds to a full "The best" strategy, while a "Mixed" strategy with $100\%$ of random promotions corresponds to a full "Random" strategy. 
\\
Finally, as already anticipated, we have also considered the possibility to promote a member from one level to the next without considering the links of the hierarchical tree, in order to reduce the new model to that one considered in ref. \cite{Peter1}: this last mode of promotion is, from here on, called {\it global mode} and refers to the old pyramidal topology, while the {\it neighbors mode} is the promotion mode which follows the links of the full modular network.

\section{New Simulation Results}
  
We present in the following the results of simulations performed with the new hierarchical model just introduced.
First of all let us try to reproduce the old  results shown in Fig.2 within this new model, focusing our attention (i) on the influence of the topology on those results and (ii) on the competition between the "Random" strategy and "The best" strategy under the Peter hypothesis of competence transmission.

\subsection{Comparison between the old and the new model}

In Fig.4 we  plot the time evolution of the relative global efficiency $E_r(\%)$ for the hierarchical tree networks showed in Fig.3, but in the 'global' mode, i.e. without considering the links: it is a situation equivalent to the old pyramidal topology showed in Fig.1, only with a different increasing size. For each network, we adopt a mixed strategy of promotions, with an increasing percentage of random promotions and coupled with the Peter hypothesis (PH). Results are averaged over $30$ events, i.e. runs with different realizations of the initial conditions. As previously mentioned, each simulation starts with a meritocratic transient of $1000$ months (not plotted in Fig.4), which is long enough to reach a stationary state for the global efficiency. Immediately {\it after} the transient the various percentages of random promotions are introduced and for the next $1000$ months we report, in each panel of Fig.4, the relative global efficiency $E_r(\%)$ for the correspondent organization, calculated as the difference between the actual absolute efficiency $E(\%)$ and the efficiency $E_{trans}(\%)$ calculated averaging along the stationary state of the transient. In other words, $E_r(\%)$ measures the gain or loss in efficiency with respect to the 'naively' meritocratic regime (whose reference value, corresponding to $E_r(\%)=0$, is indicated with a dashed line in all the plots).
\begin{figure}  
\begin{center}
\epsfig{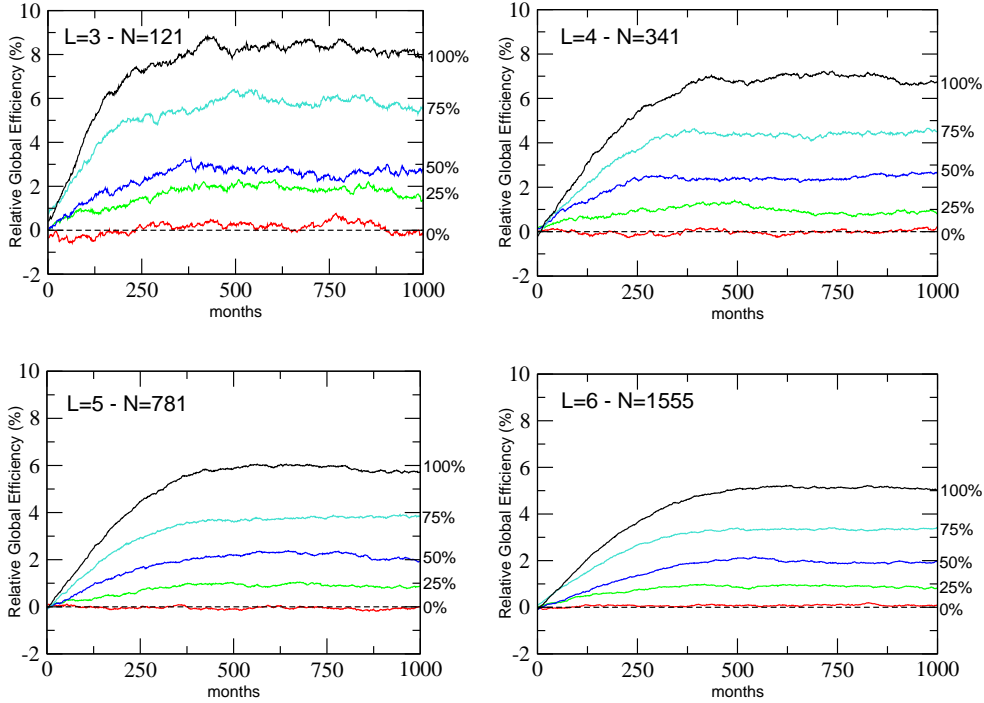}
\end{center}
\caption{ {\it Time evolution of the relative global efficiency $E_r(\%)$ for four hierarchical tree networks with increasing coordination numbers, {\it after} a meritocratic transient (not reported in the panels). An average over 30 realizations was done. For each plot, assuming the Peter hypothesis for competence transmission, a mixed strategy of promotion was adopted with an increasing percentage of random promotions with respect to "The best" strategy - reported beside the panels. The simulations were performed in 'global' mode, i.e. each  network  is reduced to a simple pyramid where an empty position at level $k$ can be filled by {\it any} agent at level $k+1$, as in the old model \cite{Peter1}.  
}}
\label{figure4}
\end{figure}
\\
We immediately see that an increase in the percentage of random promotions (reported on the right at the end of each curve) yields a gain in efficiency for all the organizations, although in general the asymptotic values of $E_r(\%)$ slightly diminish by increasing the size of the organization. Fluctuations in the time evolution are also strongly suppressed by increasing  the size of the system, since they go from values around $\pm0.3\%$ for the $L=3$ case, to values below $\pm0.1\%$ for the $L=6$ case. In any case these fluctuations do not affect the general trend of the global efficiency and therefore the conclusions of our analysis.
\\
If one focus on the difference between the $100\%$ curves (full "Random" strategies) and the $0\%$ ones (full "The best" strategies), this result confirms how previously showed in Fig.2, where the difference in efficiency (about $8\%$) between the cases "Random" + PH and "The best" + PH is comparable with that one obtained here for the $L=3$ organization, the more similar in size with the old pyramidal model. It is interesting to notice that the increase in efficiency triggered by the random promotions (also if they are present in a small percentage) is sudden and relevant immediately after the meritocratic transient, even if the system reaches its stationary states only after  twenty years ($240$ months).       
\begin{figure}  
\begin{center}
\epsfig{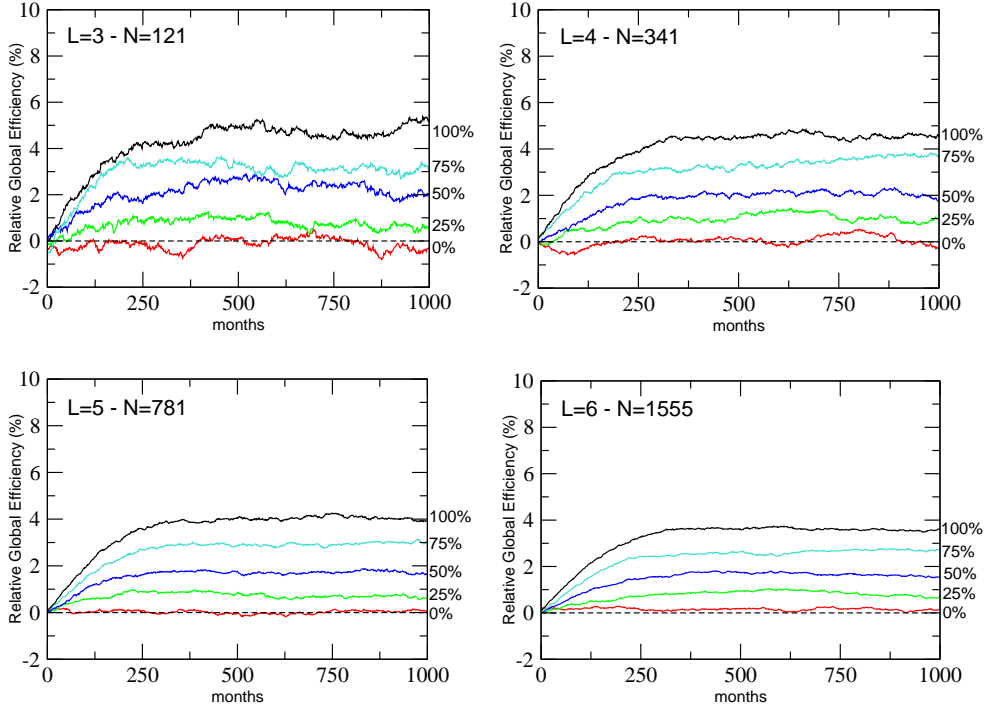}
\end{center}
\caption{ {\it We show here  simulations  analogous to those shown in Fig.4 but obtained with promotions in 'neighbors' mode, i.e. an empty position at level $k$ can be filled only from first neighbors agents at level $k+1$.    
}}
\label{figure5}
\end{figure}
\\ 
In Fig. 5 we illustrate  simulations analogous to those  presented in Fig.4, where the time evolution of the relative global efficiency $E_r(\%)$ has been calculated for the same four hierarchical tree networks shown in Fig.4, but applying the mixed strategy of promotions (again coupled with Peter hypothesis) in 'neighbors' mode, i.e. promoting  people by  following the links for climbing the hierarchy.  
The results show that the global efficiency is quite sensitive to the topology of the organization, whose effect seems to be that of shrinking the gap of gain between the full "Random" strategy efficiency ($100\%$ curve) and the meritocratic one ($0\%$ curve), for all the organization's sizes (for the $L=3$ network the gap shrinks from the $8\%$ of Fig.4 to $4\%$, even if it stays quite constant increasing $L$). This would mean that, for a given organization, a hierarchical tree topology with modules, groups and subordinates would allow to reduce the effectiveness of random strategies in mitigating the effects of Peter hypothesis with respect to the simple structure with global promotions among levels. But this is only one half of the story. In fact if one observes the absolute values of transient efficiency $E_{trans}(\%)$ (not reported in the figure), it results that, regardless of $L$, it is always greater of about $3\%$ in 'neighbors' mode with respect to 'global' mode. 
\begin{figure}  
\begin{center}
\epsfig{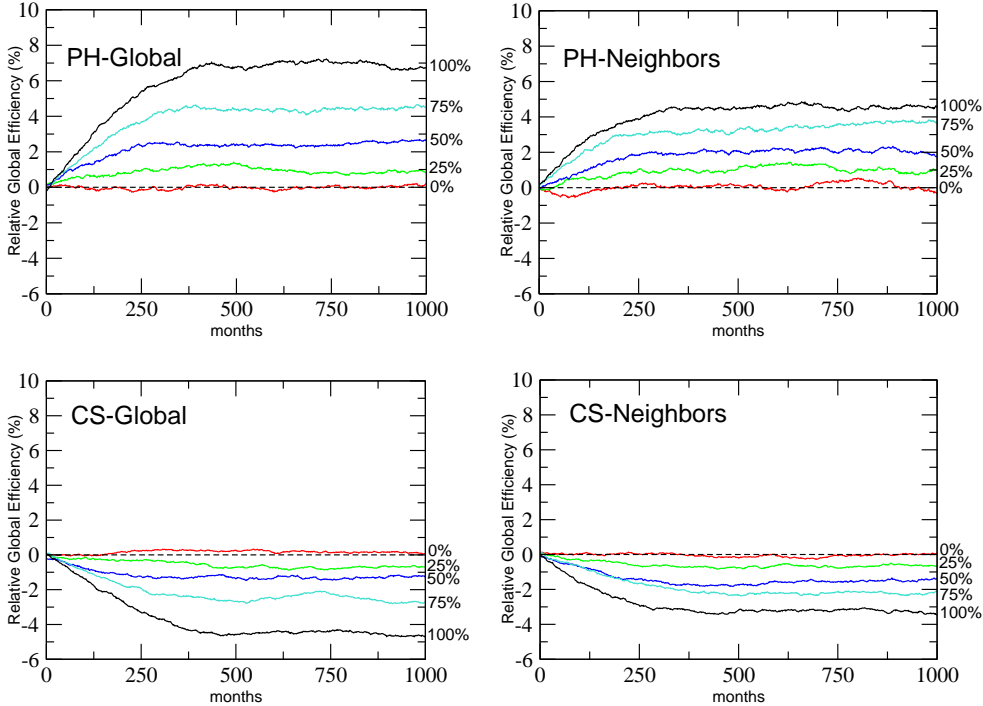}
\end{center}
\caption{ {\it Time evolution of the relative global efficiency (averaged over $30$ events) for the $L=4$ hierarchical tree network. In the first row we report, for comparison, the same plots of Fig.4 and Fig.5 for the $L=4$ network, while in the second row we show analogous plots but with the Common Sense (CS) hypothesis of competence transmission for both 'global' (left) and 'neighbors' (right) mode. As expected, in this case the introduction of a percentage of random promotions after the meritocratic transient yields a decrease in efficiency, according to the results of Fig.2, but this decrement is less in the case of the "neighbors mode", see text.   
 }}
\label{figure6}
\end{figure}
This can be explained by considering that, during the  meritocratic transient, the strategy of promoting each time the agent with the highest competence penalizes organizations with pyramidal topology  ('global' mode): in fact, in this case, the selection of the best agent at level $k$ is performed on a statistical sample with many more agents ($L^{k+1}$ to be precise) than in the case of a network topology in 'neighbors' mode (where the selection is performed only over the $L$ first neighbors), therefore the probability of promoting an agent with competence close to the maximum possible value ($10$) is very much higher. And since each promotion implies a random change of competence, as imposed by the Peter hypothesis, such a statistical effect produces, in the long term, a lower (in absolute value) transient efficiency for the pyramidal topology. On the other hand, the hierarchical tree networks in 'neighbors' mode shows a greater transient efficiency but reduces the gain in efficiency due to the introduction of random promotions after the transient, as observed in Fig.5 (we will discuss this effect with more details in the next section).  
\\
Finally, in Fig.6 we perform a last test in order to verify the consistency of the new model with the old one also when the Common Sense hypothesis for competence transmission is adopted after a  meritocratic transient, again as function of an increasing percentage of random promotions and for both the 'global' and the 'neighbors' mode. We consider here only the $L=4$ network, representing an organization with five levels and a total of $341$ members, and compare the behavior already shown in Fig.5 for this network under the Peter hypothesis (PH, top panels) with the analogous results obtained now under the Common Sense hypothesis (CS, bottom panels). It clearly appears, in agreement with the results of Fig.2 for the limiting cases of $0\%$ and $100\%$ of random promotions, that the evolution of the relative global efficiency $E_r(\%)$, averaged over $30$ events, shows a sudden decrement in the CS regime when even a small percentage of randomness is introduced in the promotion strategy. This confirms the effectiveness of promoting the best members when the new competence requested at level $k$ is correlated with the old one requested at level $k+1$. In this case the modular structure of the organization ('neighbors' mode, right bottom panel) reduces the negative effects with respect to the pyramidal one ('global' mode, left bottom panel), therefore a complex topology with project teams, managers and subordinates, is again recommended for a real hierarchical organization also if Peter hypothesis does not hold.

\subsection{Robustness of the random strategy gain}

In order to better compare the total efficiency gain which takes into account the contributions of both the topology and the random promotions, it is more convenient to represent the results of the upper panels of Fig.6, concerning a $L=4$ network, in the form of a histogram, as shown in Fig.7. Here the global efficiency in the 'global' mode (on the left) corresponds to the asymptotic efficiency gain due to the introduction of random promotions after the meritocratic transient, while the global efficiency in the 'neighbors' mode (on the right) is calculated by taking as reference value the average transient efficiency $E_{Gtrans}(\%)$ in the 'global' mode (put equal to $0$), and by adding the gain in efficiency $E_{Ntrans}(\%)-E_{Gtrans}(\%)$, evaluated at the end of the transient and due to the modular topology (a gain which depends on the statistical effect previously explained and stays here around $3\%$), to the asymptotic gain in efficiency due to random promotions. Fluctuations around the average stationary values are not visible in this representation, but they remain very small and of the same order of those of the previous figures for the case$L=4$. 
\\
Within this new visualization it becomes evident that the modular complex structure ('neighbors' mode) is more convenient for a real organization (in terms of efficiency) not only when the Peter hypothesis (PH) holds and one adopts the strategy of promoting the best members (as happens during the transient), but also when one adopts a percentage of random promotions less than $100\%$ (again under the PH). Only applying the full "Random" strategy the global efficiency seems to no longer depend on the topology. 
With this last figure as reference, let us now  test the robustness of the random strategy gain with respect to the introduction of new realistic features to our model. Of course we will introduce one feature at the time keeping fixed all the others, in order to better emphasize the influence on the organization  for a given modification. 

\begin{figure}  
\begin{center}
\epsfig{figure=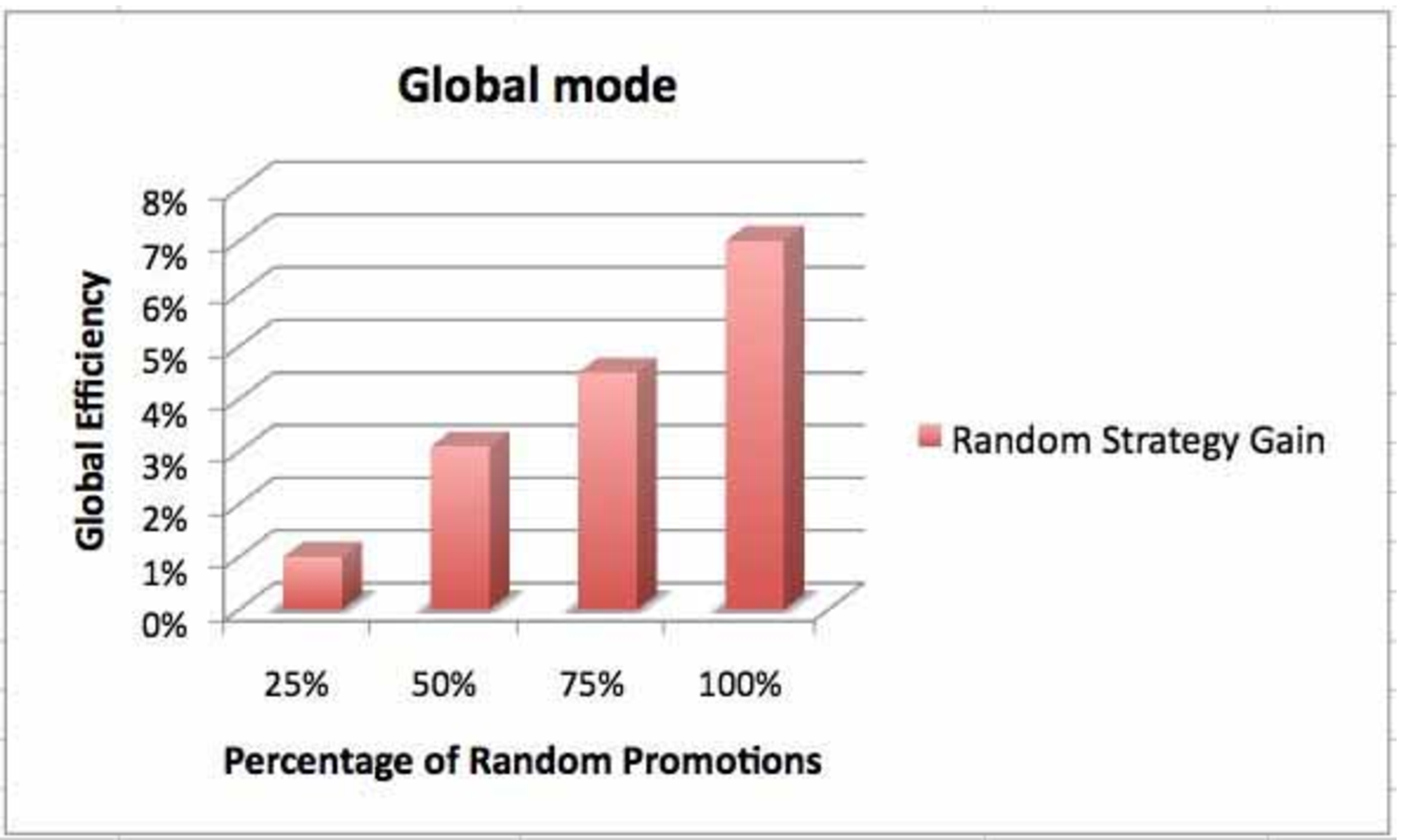,width=6.8truecm,angle=0}
\epsfig{figure=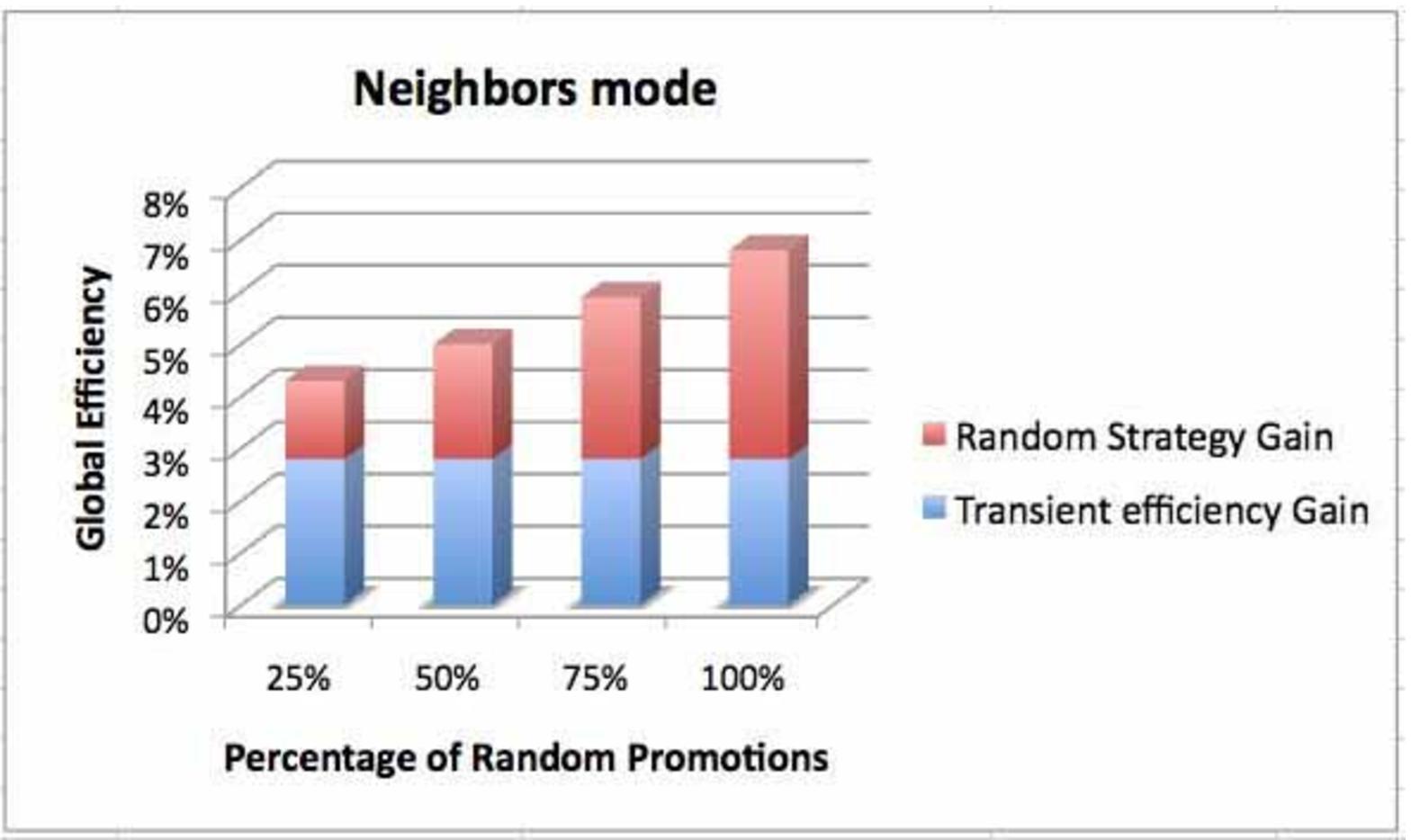,width=6.8truecm,angle=0}
\end{center}
\caption{ {\it Global efficiency gain for the $L=4$ hierarchical tree network for an increasing percentage of random promotions in both 'global' and 'neighbors' mode. The different contribution due to the transient dynamics, with meritocratic promotions, and to the introduction of random strategies, is emphasized (see text).     
}}
\label{figure7}
\end{figure}  

{\it Simulations with age-dependent competences}

In the previous section, as well as in our previous paper \cite{Peter1}, the competence of the various agents was considered fixed in time at the top of their respective possibility. But in a more realistic situation one can imagine that the competence of a young employee would improve until a certain age and then slowly diminishes until retirement.
Furthermore, very often employees are frustrated by a lack of promotion opportunities, and individual efficiency decreases when an increase in age does not match with a proper career progression: such an effect is known as the {\it Prince Charles syndrome} and has been already studied also computationally \cite{Klimek1}.
\\
 \begin{figure}  
\begin{center}
\epsfig{figure=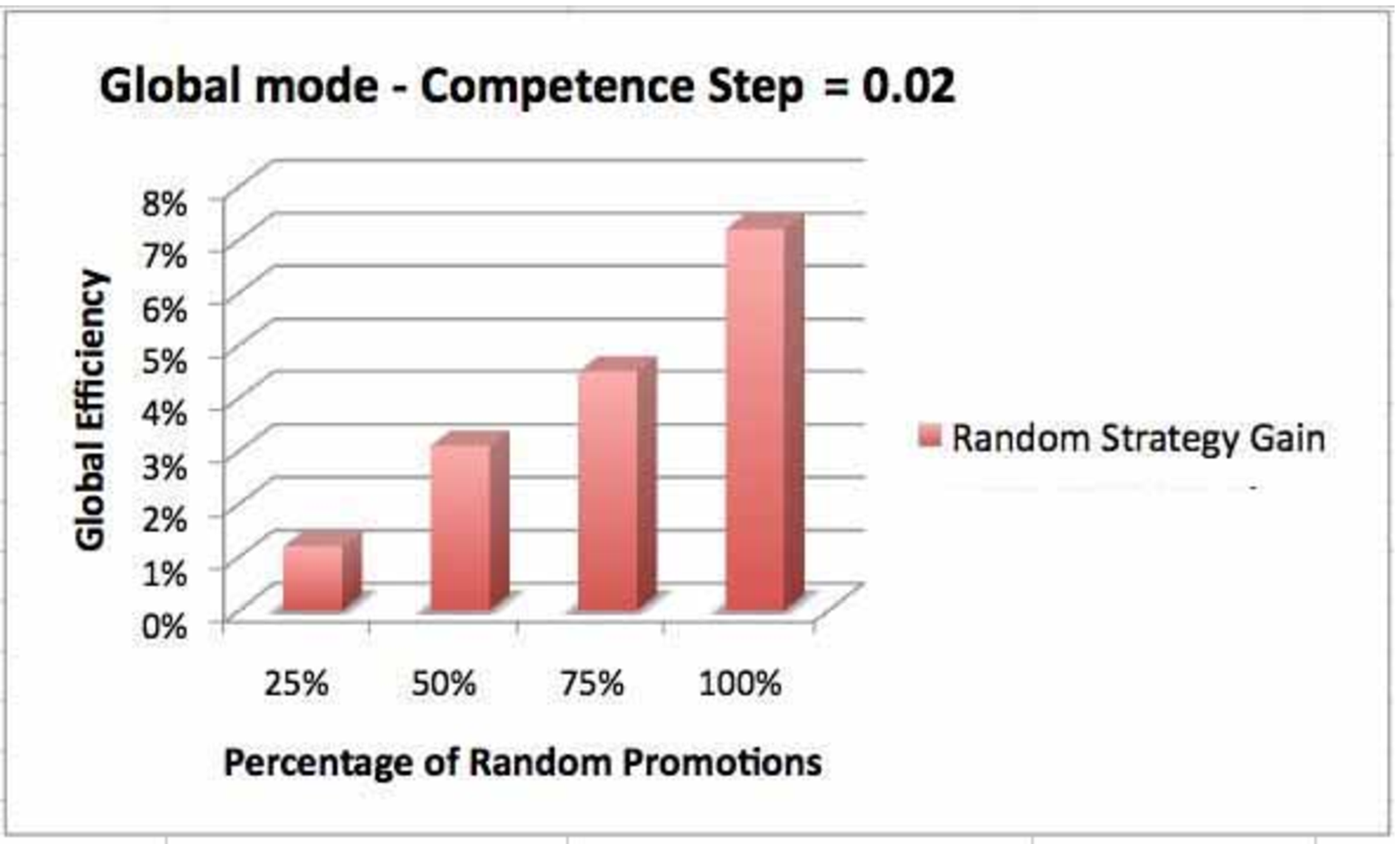,width=6.8truecm,angle=0}
\epsfig{figure=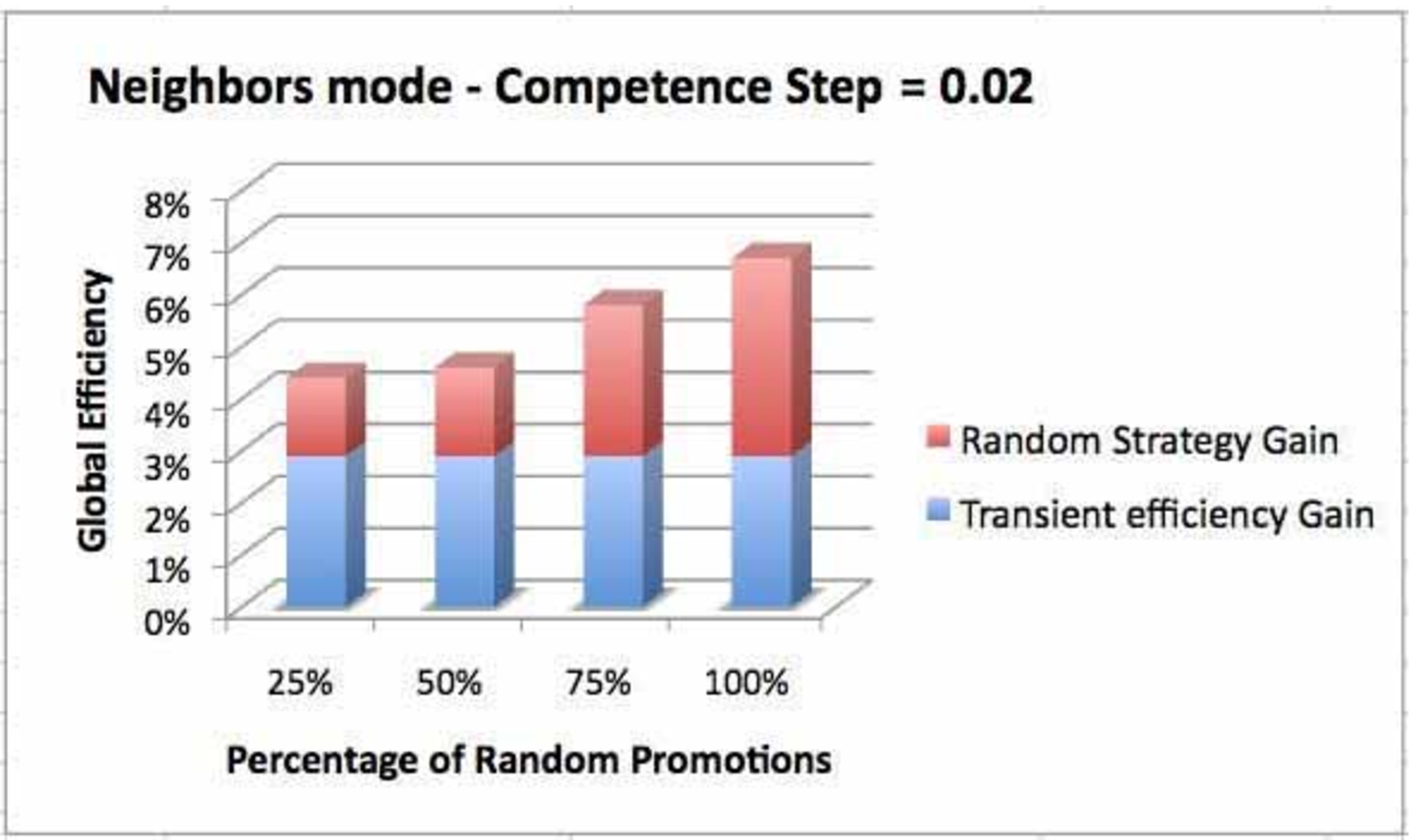,width=6.8truecm,angle=0}
\epsfig{figure=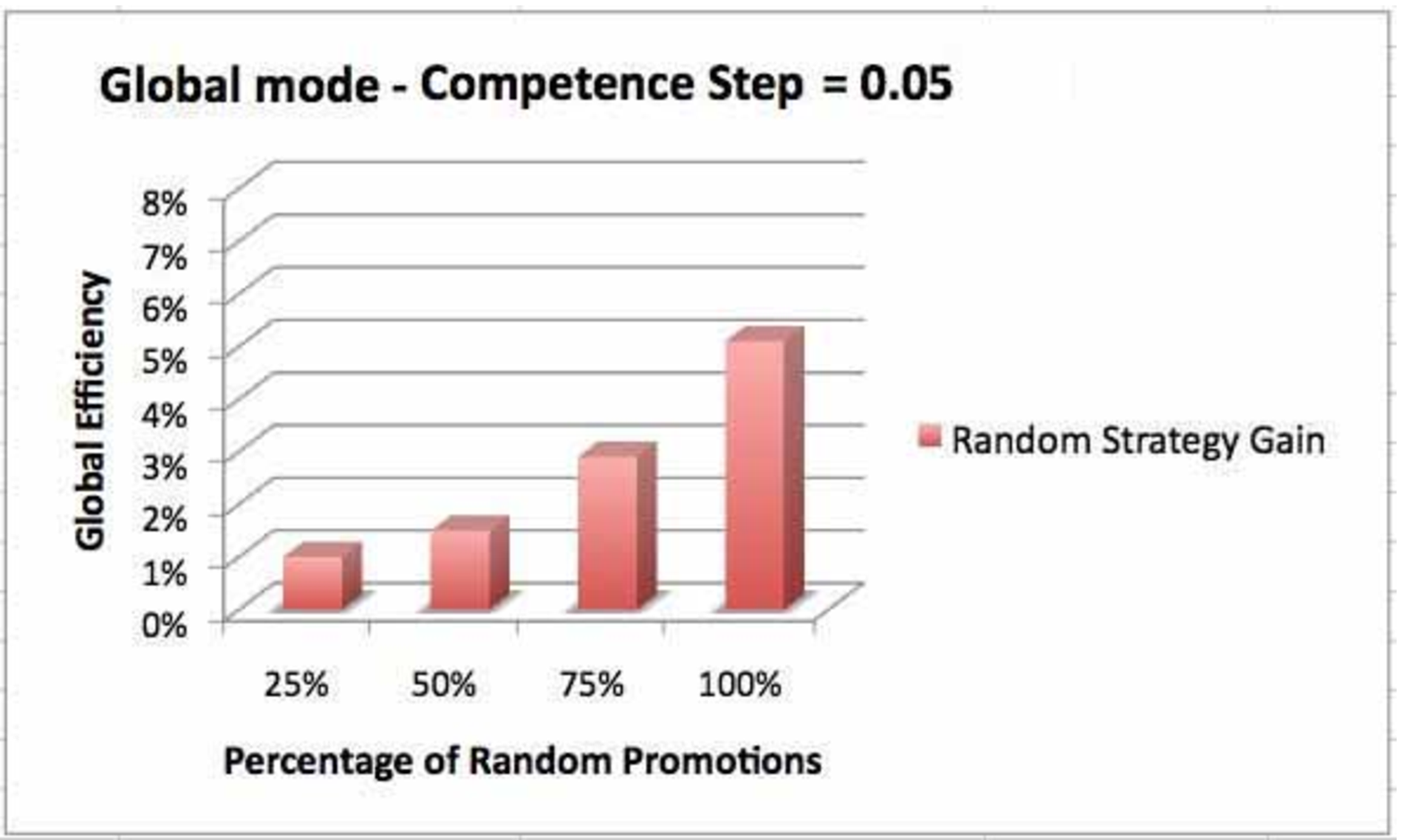,width=6.8truecm,angle=0}
\epsfig{figure=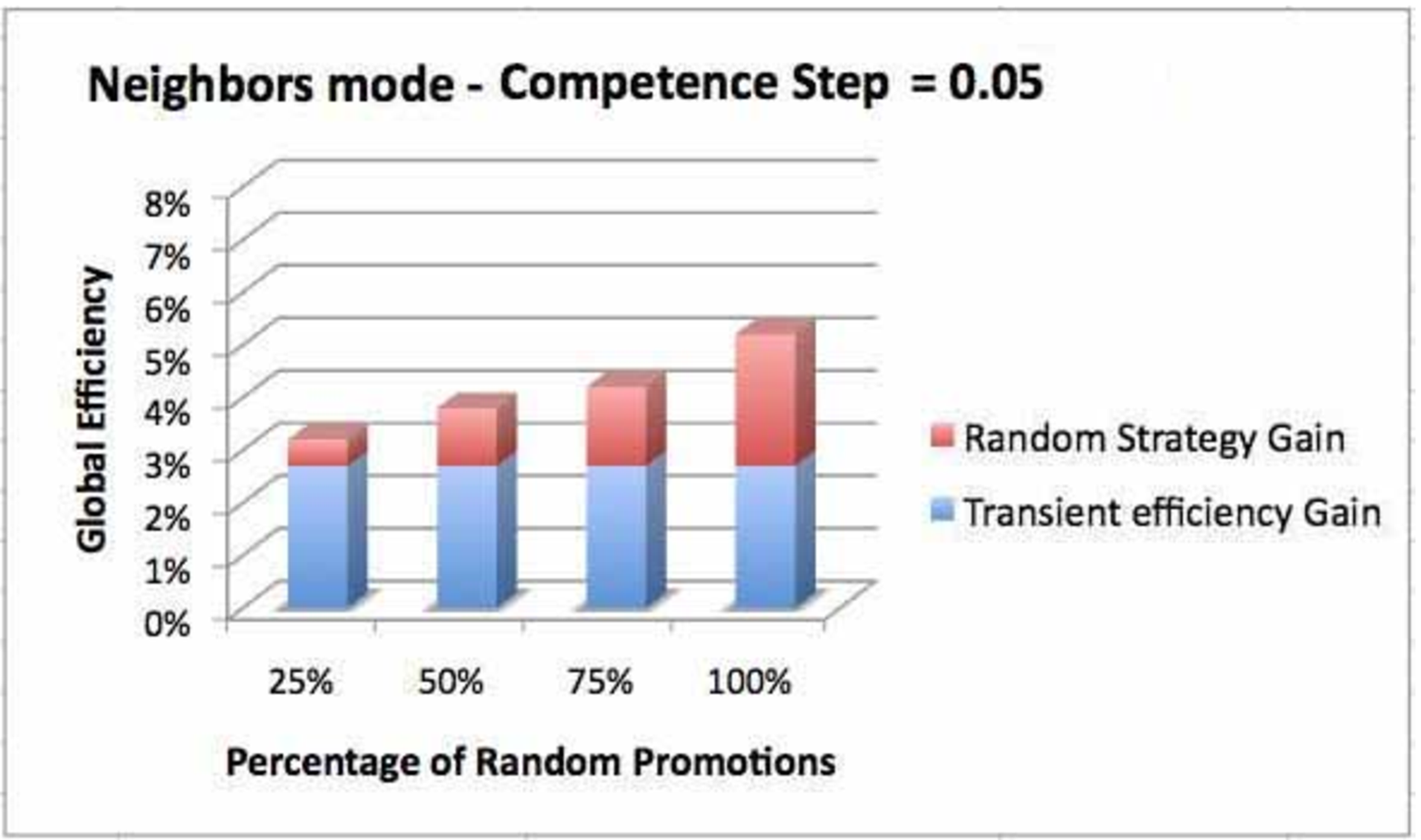,width=6.8truecm,angle=0}
\end{center}
\caption{ {\it Global efficiency gain (averaged over 30 events) for the $L=4$ hierarchical tree network for an increasing percentage of random promotions in both 'global' (left side) and 'neighbors' (right side) mode. In these simulations we introduced an annual increment of competence (+0.02 and +0.05) for each agent during the first part of his career (independently from the level) up to an age of $38$ years, and an annual decrement of competence ($-0.02/r_k$ and $-0.05/r_k$) during the remaining part of career (see text for more details).     
}}
\label{figure8}
\end{figure}  
It is therefore interesting to explore if the effectiveness of the random promotion strategy holds when a competence variable in time with the age and the position of the agents is introduced in our new model. 
In particular, we introduced an increment of competence in time for each agent during the first part of his career in the organization, independently from the level, up to an age of $38$ years, and a decrement of competence during the remaining part of career which, however, further decreases climbing the hierarchy in order to take into account the Prince Charles syndrome (at the same age, the higher is the level, the greater is the motivation to carry out own work with efficiency). In Fig.8 we plot the global efficiency gain for a $L=4$ network, in both 'global' and 'neighbors' mode, as function of an increasing percentage of random promotions (under the PH) and for two different steps of increment/decrement of competence. In the two upper panels we fix an annual competence increment of $+0.02$ and an annual decrement of $-0.02/r_k$, being $r_k$ the responsibility of level $k$ (which increases linearly with the levels, as reported in Fig.1), while in the lower panels the annual increment is $+0.05$ and the annual decrement $-0.05/r_k$: in both the cases, from a comparison with Fig.7, it clearly appears that this new feature does not affect sensitively the global efficiency, independently of the topology. Actually, apart from a small general loss of $\sim2\%$ observed in the lower plots, the gain induced by random promotions tends to be confirmed for both the pyramidal and the modular structures of the organization.    

\begin{figure}  
\begin{center}
\epsfig{figure=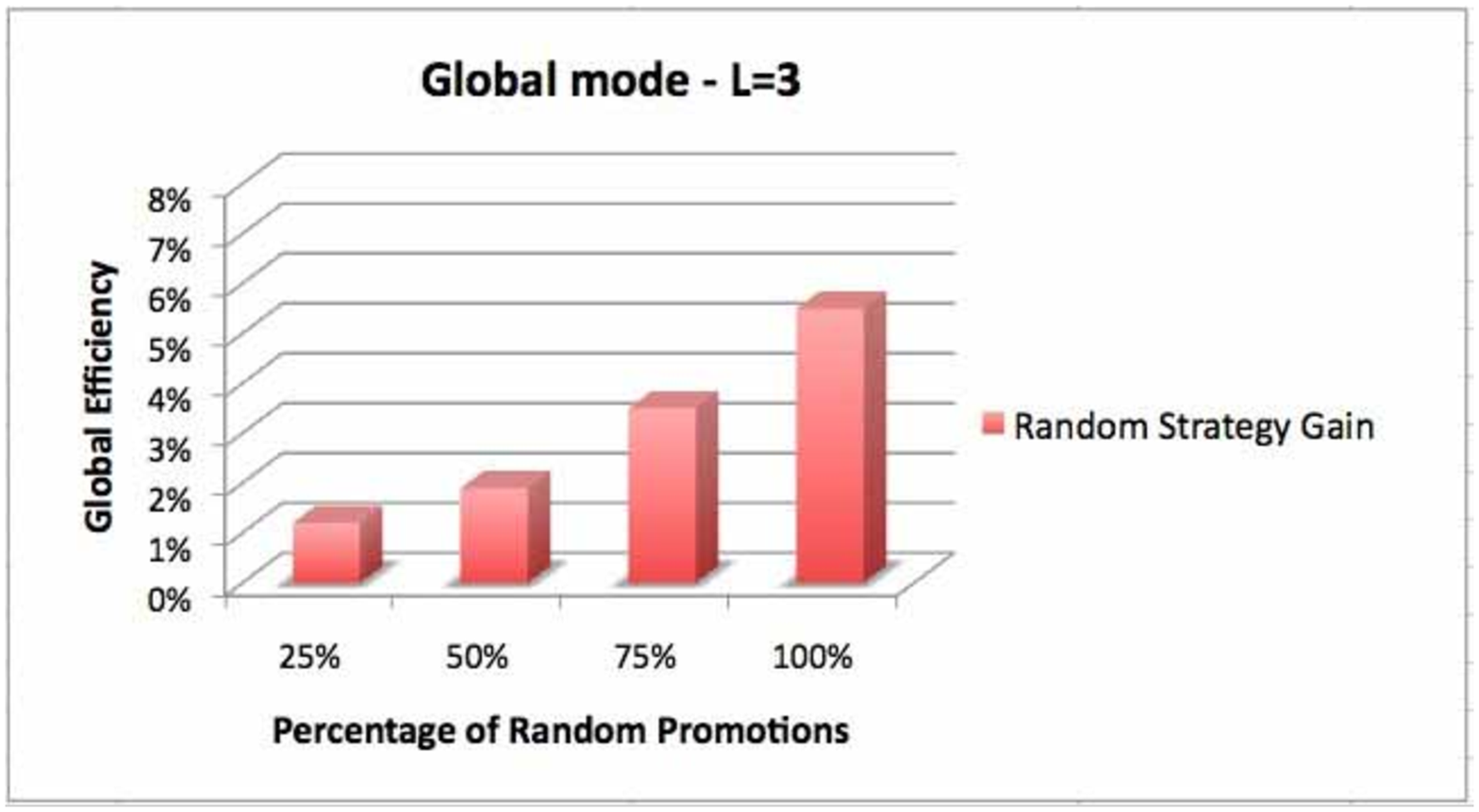,width=6.8truecm,angle=0}
\epsfig{figure=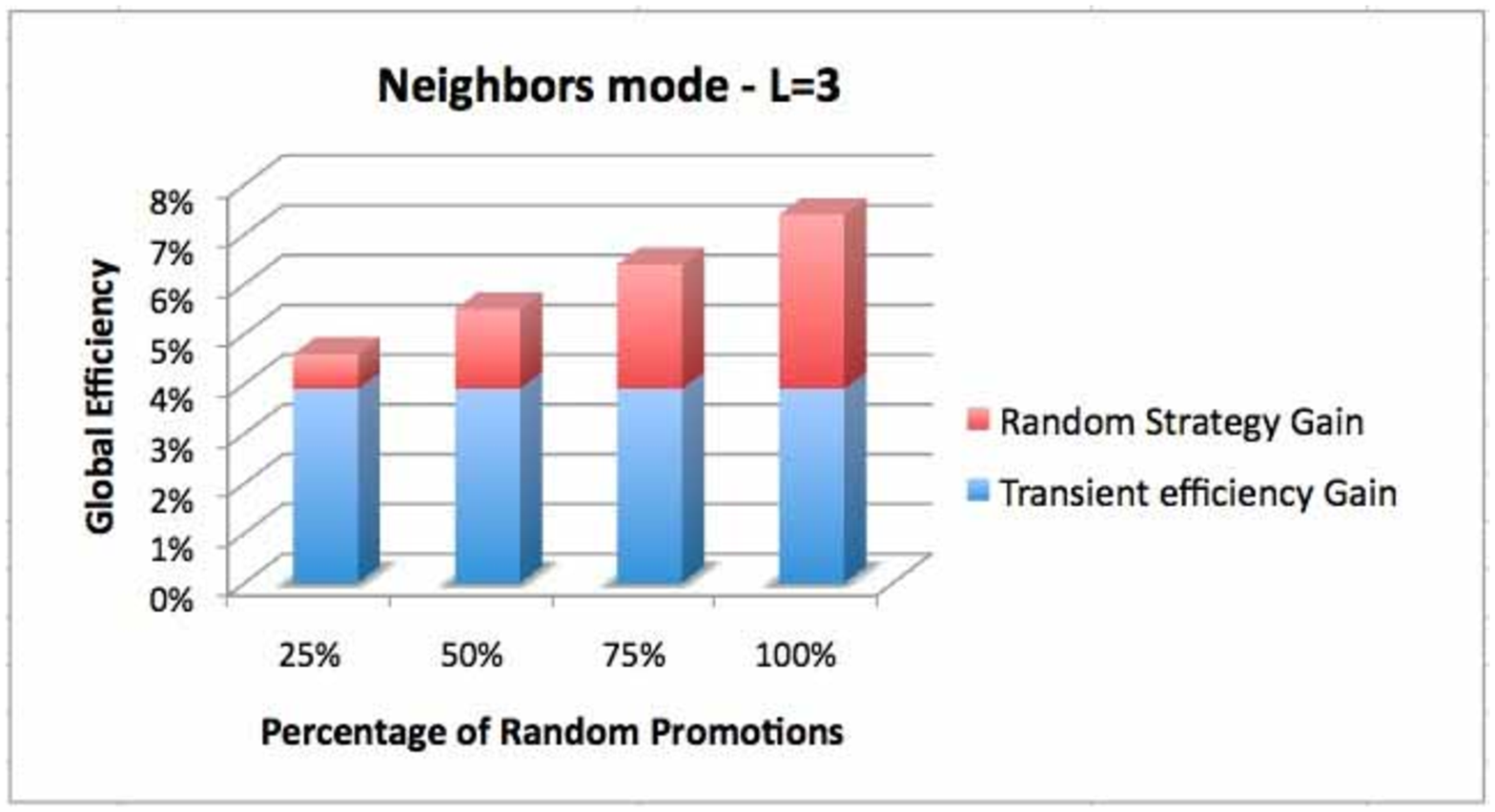,width=6.8truecm,angle=0}
\epsfig{figure=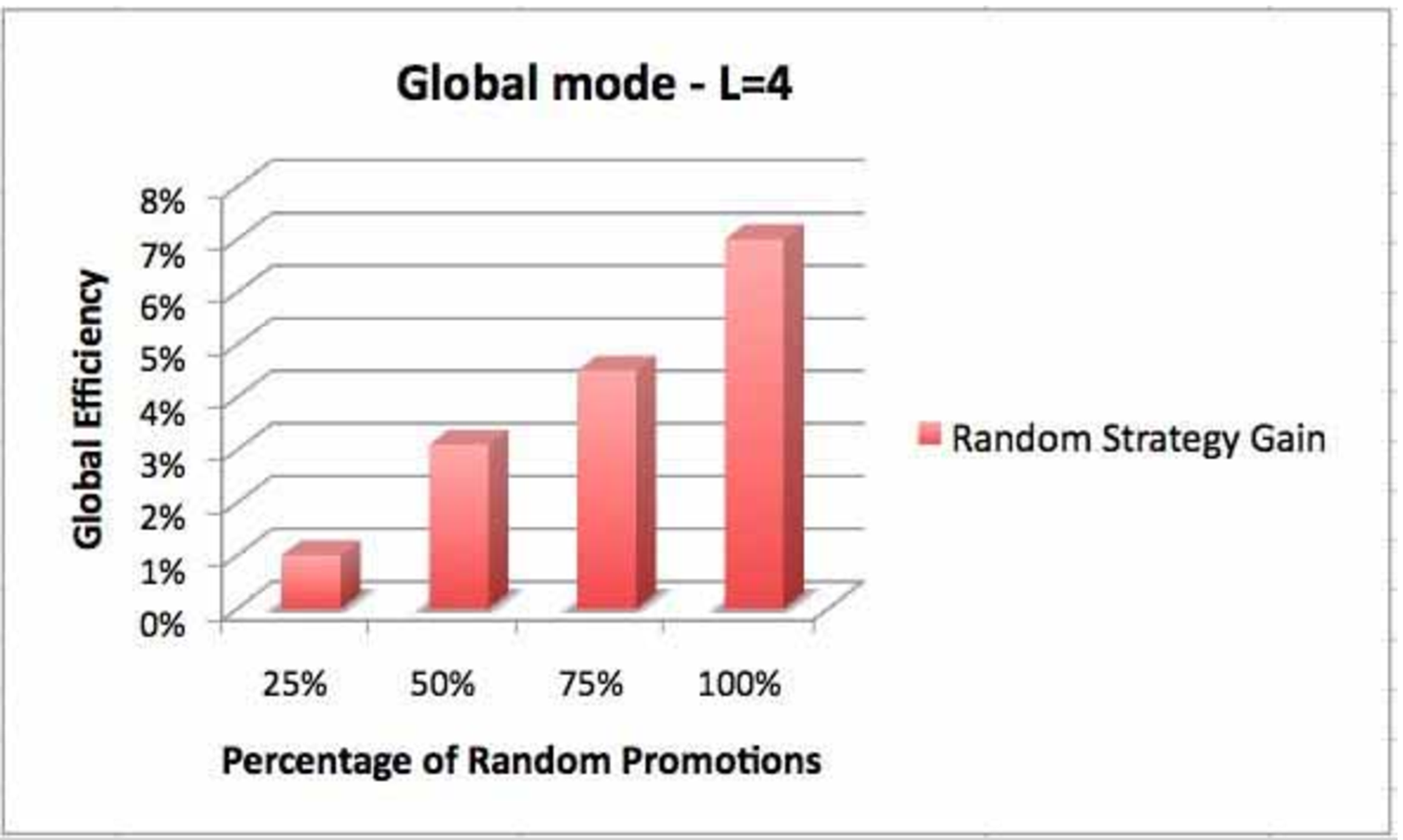,width=6.8truecm,angle=0}
\epsfig{figure=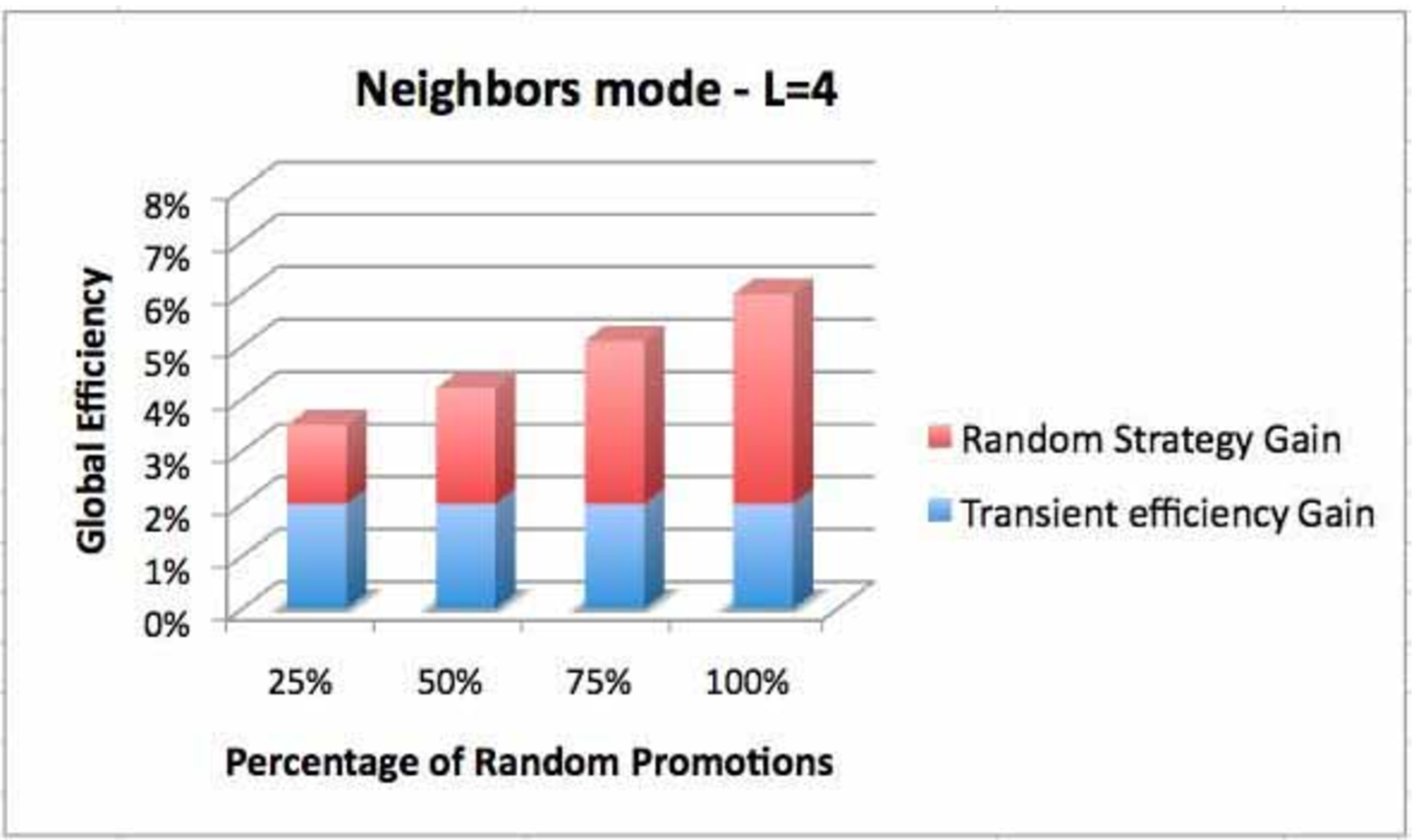,width=6.8truecm,angle=0}\end{center}
\caption{ {\it Global efficiency for the $L=3$ and $L=4$ 
hierarchical tree networks (averaged, respectively, over 30 and 20 events), for an 
increasing percentage of random promotions in both 'global' 
and 'neighbors' mode of Peter hypothesis. In order to make more realistic the responsibility increment from 
the bottom to the top level, in these simulations we adopted, instead of the usual linear scale, a non linear one, see text for more details.
}}
\label{figure9}
\end{figure}

{\it Organizations with nonlinear responsibility}

In all the previous simulations, as already observed, the responsibility of the five levels of the organization varies linearly from $0.2$ to $1$, but of course this is an arbitrary criterium and one could ask how the response of the system to the introduction of random promotions could be affected by changing it. Actually, since the responsibility of managers within a hierarchical organization could  be likely considered correlated with the total number of their subordinates at the lower levels, it is probably more realistic to adopt a non linear scale for the responsibility. In this respect we did some simulations in which  the responsibility $r_k$ of an agent at level $k<5$ is equal to the total number of his/her subordinates at all the levels below, i.e. $N_k=L^{6-k}-L/L-1$. For $k=5$ we put $r_k=1$. 
The results  are reported in Fig.9 for a $L=3$ and a $L=4$ organizations, in both 'global' and 'neighbors' modes and under the Peter hypothesis of competence transmission. 
Again, also in this case, we observe a positive reaction of the asymptotic global efficiency to the adoption of random strategies, with a total gain that is independent of the topology only for full random promotions, while for a percentage of randomness smaller than $100\%$ the contribute of topology becomes important and privileges the 'neighbors' mode, i.e. a modular structure with respect to a pyramidal one. Anyway, comparing the lower panels of Fig.9 with those of Fig.7, apart of a slight decrease ($\sim-1\%$) in transient efficiency for the 'neighbors' mode, results are very similar, therefore the choice of the responsibility scale does not appear to affect so much the global efficiency of the system and the gain due to random promotions. 

\begin{figure}  
\begin{center}
\epsfig{figure=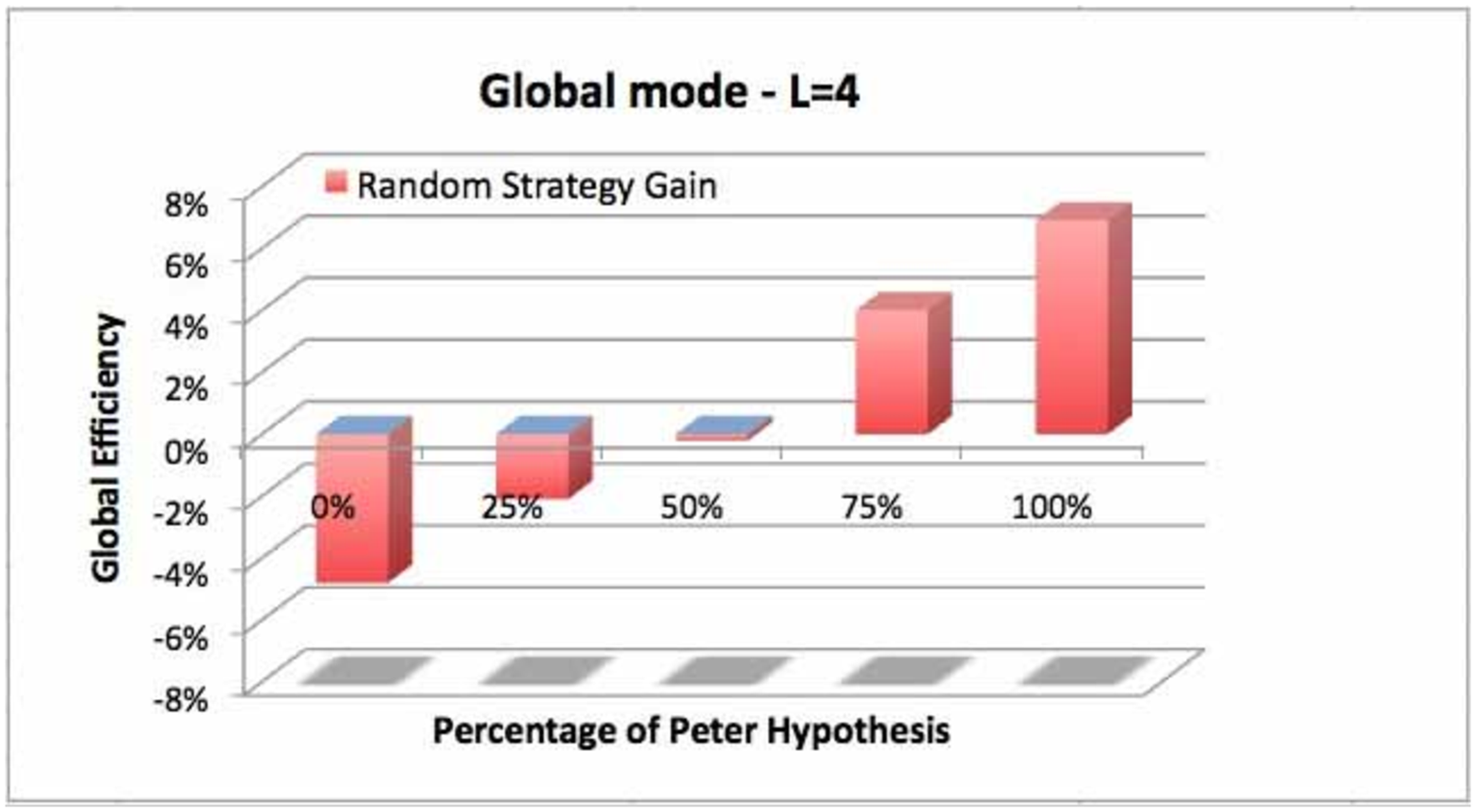,width=9truecm,angle=0}
\epsfig{figure=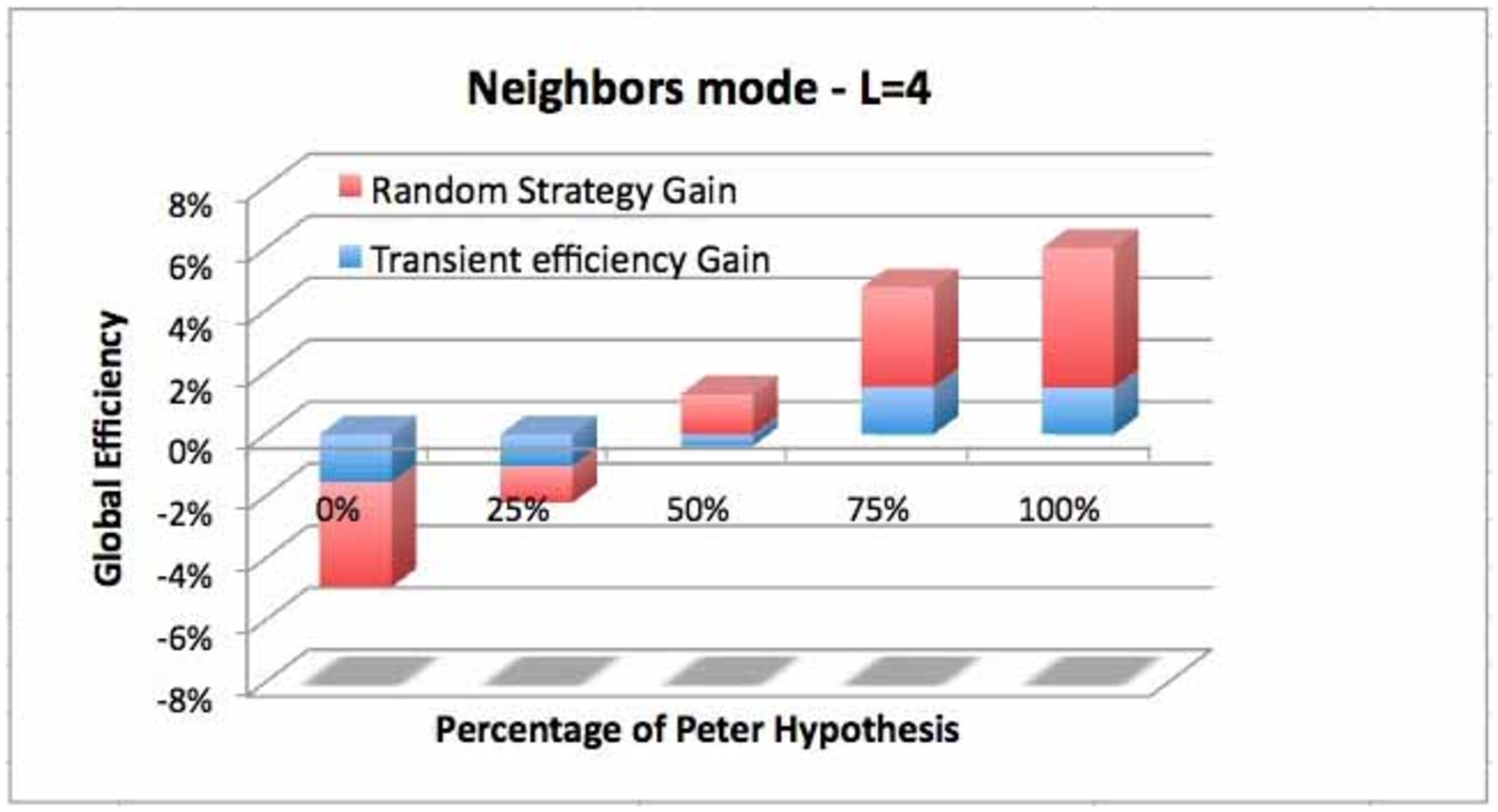,width=9truecm,angle=0}
\end{center}
\caption{ {\it Global efficiency (averaged over 30 events) for the $L=4$ hierarchical tree network in both 'global' and 'neighbors' mode and with a full ($100\%$) random strategy. In these cases we simulate an organization where a certain percentage of positions are characterized by the Peter hypothesis (indicating no correlation between competences before and after the promotion at those positions), while the remaining ones are characterized by the Common Sense hypothesis (indicating a strong correlation between competences before and after the promotion at those position). See text for further details.}}
\label{figure10}
\end{figure}

{\it Mixed hypothesis for the transmission of competences}

Let us now  consider a situation that is probably very common in real organizations, where often neither the Common Sense nor the Peter hypothesis of competence transmission are always satisfied, individually, for {\it all} the positions. Rather, we can suppose that, in many real cases, some particular positions in the hierarchy will require, for  the promoted agent, a small change in the task to perform with respect to the previous level (CS), while other positions could require very different skills (PH). In order to take into account also these situations, in the next simulations we consider the possibility of having two different kinds of positions randomly distributed over the organization, one satisfying the PH and the other one the CS, but with a different probability. 
\\
In Fig.10 the behavior of the asymptotic global efficiency is reported for the usual $L=4$ organization (with linear responsibility scale) in both 'global' and 'neighbors' mode, now as function of an increasing percentage of PH positions. As always, each simulation starts with a meritocratic transient characterized by the promotion of the best members, after which we apply here a full ($100\%$) random strategy of promotions. Notice that, at variance with the previous simulations, here both the $E_{Ntrans}(\%)$ and $E_{Gtrans}(\%)$, and then the gain in efficiency $E_{Ntrans}(\%)-E_{Gtrans}(\%)$ due to the topology, may change depending on the percentage of PH positions, since the latter is a structural feature which affects also the transient dynamics. Furthermore,  $E_{Ntrans}(\%)-E_{Gtrans}(\%)$ can now also be negative when the percentage of PH positions is lower than $50\%$, since in that case the Common Sense hypothesis prevails in the organization and the meritocratic transient benefits the 'global' pyramidal topology with respect to the 'neighbors' one (for the same statistical effect which favors the modular structure when the Peter hypothesis holds). 
\\
From Fig.10 we see that, for any topology, the application of the $100\%$ random promotion strategy produces positive effects only for organizations with more than $50\%$ of PH positions, while in the other cases promoting the best members continues to be a winning strategy. This definitively clarifies that adopting random promotions is not {\it always} a recommended strategy, but {\it only} when in a given organization the Peter hypothesis holds for the majority of the positions.              

\begin{figure}  
\begin{center}
\epsfig{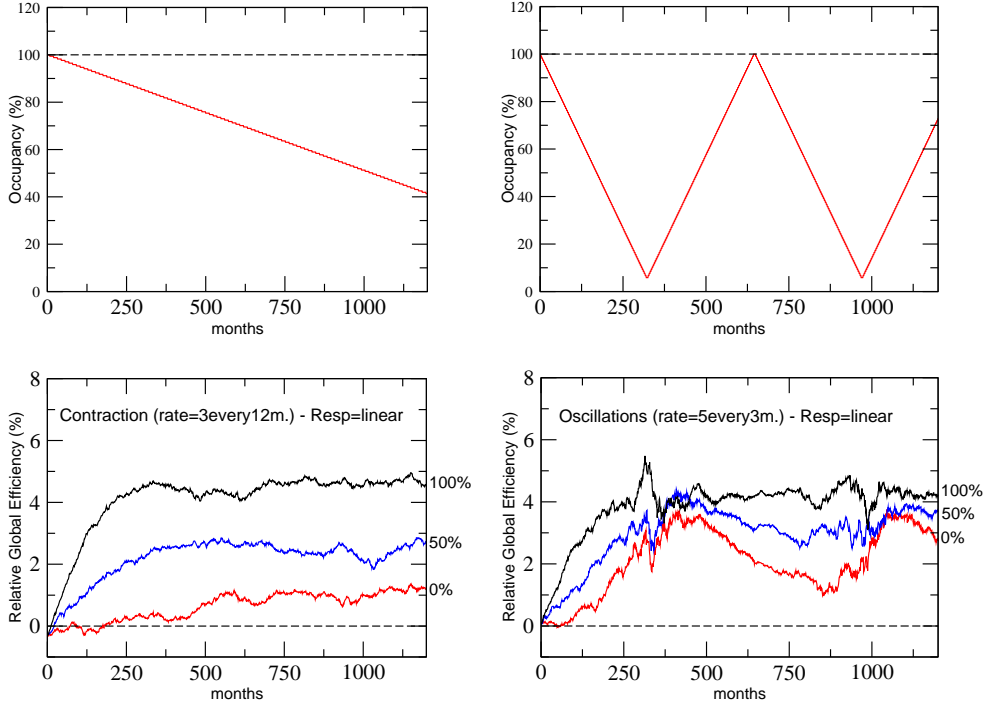}
\end{center}
\caption{{\it In the bottom panels we show the time evolution of the relative global efficiency for the $L=4$ hierarchical tree network (averaged over 30 events), for an increasing percentage of random promotions in 'neighbors' mode of Peter hypothesis. The novelty is the introduction of the possibility, for the organization, to change in time its number of active positions. A contracting and an oscillating organizations have been considered, respectively, on the left and on the right panels. Fluctuations in efficiency are visible especially in the right panel, due to the sudden inversion of tendency in the size oscillation.       
}}
\label{figure11}
\end{figure}

{\it Time-dependent size of the organization}

Our last test concerns the introduction of the possibility  to change in time the size of  the organization. Actually, in the previous simulations, the total number of active positions in the organizations was kept fixed in time, but this constraint could be considered not very realistic since companies can decrease or increase the number of their employees in time following periods of crisis or expansion. Therefore we tried to explore the effects of the introduction of a given percentage of random promotions in a contracting or in an oscillating modular $L=4$ organization (in 'neighbors' mode, under the Peter hypothesis) after a meritocratic transient during which the size is held constant in time. The results of the simulations are shown in Fig.11.
First of all we let the organization contract in time by dismissing its members with a rate of $3$ positions every $12$ months, as shown in the left top panel, where the percentage of active positions (i.e. te level of occupancy) is plotted as function of time; correspondently, in the panel below we plot the behavior of the relative global efficiency for three different percentages of random promotions. The introduction of even a small percentage of random promotions quickly enhances the relative efficiency also during a period of crisis.          
The same behavior can be found in the plots on the right side of Fig.1, where we let the number of active positions of the organization oscillate in time with a dismissal rate of $5$ positions every $3$ months, as shown in the top panel. Again, random strategies (and in particular the full $100\%$ one) seem to be still more effective than the others, as shown in the panel below. In general we also see that the effect of reducing the size improves the efficiency while an expansion reduces it, until, with an occupancy under the $40\%$, even the meritocratic strategy with $0\%$ random promotions improves the efficiency of the organization (but, in any case, less than the random one). Therefore also 
 these simulations further confirm the robustness of the full random strategy, which can be definitively considered as the more effective for any complex organization, pyramidal or modular, for which the Peter hypothesis applies.

\section{General discussion }

In this section we will try to summarize the results of our simulations presented  previously  and discuss the possible application in real cases.
In all the examples we have presented, a common feature strongly emerges: the efficiency of an organization increases significantly  if  one adopts a random strategy of promotion with respect to a simple meritocratic promotion of the best members. This fact, already shown in our previous paper \cite{Peter1} for a very simple pyramidal model and under a minimum number of assumptions, has proven to be very robust and persistent even in a new hierarchical tree model and under many different kinds of realistic improvements. 
\\
Actually, the finding  of the effectiveness of random promotions is not completely new. In fact,  we have recently discovered  that some years ago two researchers of the Dallas School of Management, Texas (US), published a study of management in real companies through computer simulations and  found that promotion of best performers may actually degrade the overall organizational performance when compared with just promoting a random member of the group \cite{Phelan-Lin}. But, since they introduced random promotions as a baseline control system, they were very surprised of this unexpected result, which in any case  was not connected at all to  the Peter principle and, as far as we know, they did not returned on the subject with further investigations.
On the contrary, in our simulations the random strategy has been introduced from the beginning as a real alternative strategy to the meritocratic one, and the increase of efficiency triggered by random promotions has been identified as an emergent feature which comes out because of the cooperative effect of many promotion events under the Peter hypothesis. In this respect, our new results corroborate the fact that one does not need a full random strategy to obtain an increase of efficiency:  in many cases a choice of only $50\% $ of agents  selected in a random way for promotion results to be enough to obtain a consistent increment in the efficiency. Furthermore, in all cases discussed the random strategy improves consistently the efficiency of the system revealing a very persistent  robustness. 

\subsection{Considerations about the first $20$ years}

The simulations presented in the previous section were performed for a period of $1000$ months (about $83$ years). This period could seem too long to be considered realistic also for a real organization which wants to establish a long-term strategy of promotions. But, on the other hand, the improvement in efficiency induced by the random strategy after the transient has been shown to be rapid and substantial  since the very beginning (1\% after 3 years, 2\% after 6 years and so on) and usually the global efficiency reaches a stationary state just after  $20$ years. In order to discuss in deeper detail this point we emphasize in Fig.12 the initial part of the efficiency time evolution for a hierarchical organization with $K=5$, $L=4$ and $N=341$, considering an increasing percentage of random promotions for both a pyramidal ('global') and modular ('neighbors') topology under the Peter hypothesis of competence transmission. In particular, we focus on a period of $240$ months ($20$ years) and we also perform an average over $30$ different realizations in order to diminish the effect of fluctuations. 
\begin{figure}  
\begin{center}
\epsfig{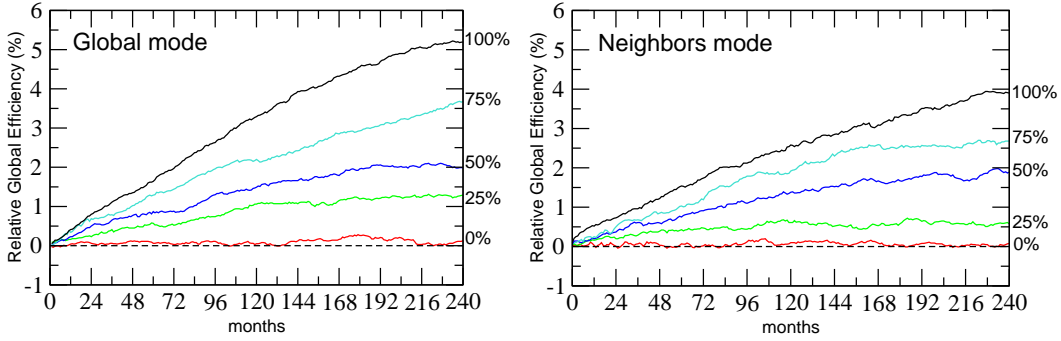}
\end{center}
\caption{ {\it A magnification of the initial evolution of Figs.4,5 that shows  an immediate increase of the efficiency  since the beginning of the adoption of a random strategy with respect to a meritocratic one with the Peter hypothesis, see text for further details.
}}
\label{figure12}
\end{figure}
The simulations strongly indicate  that, independently of the topology and of the percentage of random promotions, in the first $20$ years the global efficiency reaches values which are equal to about the $75\%-80\%$ of the asymptotic values shown in the upper panels of Fig.6. In Tables 1 and 2  we report detailed information concerning all the relevant quantities which characterize the organization considered in Fig.12 during the $20$ years. 
\\
In particular, in Table 1 we report (from left to right), for both 'global' (top) and 'neighbors'  (bottom) modes  and for each percentage of random promotions, the following quantities: (i) the gain in efficiency accumulated during the meritocratic transient (calculated with respect to the 'global' mode, which therefore has gain equal to $0\%$); (ii) the further gain in efficiency due to the adoption of a random strategy after the transient (correspondent to the maximum values reached in Fig.12); (iii) the total number of dismissals ($N_d$); (iv) the total number of retirements ($N_r$); (v) the total number of promotions ($N_p$). 
\\
On the other hand, in Table 2 we report, again for both the topologies and for all the percentages of random promotions, the total number of agents which terminated their career (due to dismissal or retirement) at each one of the $5$ levels of the organization, i.e., respectively (from left to right), $N_5$, $N_4$, $N_3$, $N_2$, $N_1$. 
\\
\begin{table} 
\begin{center}
\begin{tabular}{c c c c c c c}
\hline\hline
   {\it Global Mode - $\%$ of rnd prom.} & {\it Trans.Eff.Gain} & {\it Rnd.Eff.Gain} & {\it  $N_d$} & {\it $N_r$} & {\it $N_p$} \\
\hline
{\bf $0\%$}     & $0.00\%$ & $0.36\%$ & $22$ & $206$ & $84$ \\
{\bf $25\%$}   & $0.00\%$ & $1.01\%$ & $22$ & $205$ & $94$ \\
{\bf $50\%$}   & $0.00\%$ & $2.37\%$ & $20$ & $206$ & $96$ \\
{\bf $75\%$}   & $0.00\%$ & $3.80\%$ & $24$ & $209$ & $111$ \\
{\bf $100\%$} & $0.00\%$ & $5.25\%$ & $23$ & $204$ & $116$ \\
\hline\hline
\end{tabular}
\begin{tabular}{c c c c c c c}
\hline\hline
   {\it Neigh. Mode - $\%$ of rnd prom.} & {\it Trans.Eff.Gain} & {\it Rnd.Eff.Gain} & {\it  $N_d$} & {\it $N_r$} & {\it $N_p$} \\
\hline
{\bf $0\%$}     & $2.70\%$ & $0.14\%$ & $22$ & $203$ & $135$ \\
{\bf $25\%$}   & $2.70\%$ & $1.12\%$ & $24$ & $204$ & $136$ \\
{\bf $50\%$}   & $2.70\%$ & $1.42\%$ & $23$ & $204$ & $138$ \\
{\bf $75\%$}   & $2.70\%$ & $2.83\%$ & $25$ & $201$ & $144$ \\
{\bf $100\%$} & $2.70\%$ & $3.70\%$ & $25$ & $202$ & $146$ \\
\hline\hline
\end{tabular}
\caption{{\it With reference to the simulations plotted in Fig.12 we report, from left to right, the  percentage of random promotions in 'global' mode (upper panel) and in 'neighbors' mode (lower panel); the gain in efficiency accumulated during to the meritocratic transient (calculated with respect to the 'global' mode);  the further gain in efficiency due to the adoption of a random strategy after the transient; the total number of dismissals; the total number of retirements; the total number of promotions. See text for further details. }}
\end{center}
\label{summary-table1}
\end{table}
Looking at these Tables we immediately notice something strange. In fact, from the last three columns of Table 1 one sees that, while the total number of dismissals and retirements in the $20$ years is almost the same independently of the topology, the number of promotions is quite different in the two cases and it results to be always greater in the 'neighbors' mode (i.e. for a modular organization) of about 30 units on average.  
\begin{table} 
\begin{center}
\begin{tabular}{c c c c c c c}
\hline\hline
   {\it Global Mode - $\%$ of rnd prom.} & {\it  $N_5$} & {\it  $N_4$} & {\it  $N_3$} & {\it $N_2$} & {\it $N_1$} \\
\hline
{\bf $0\%$}     & $167$ & $44$ & $12$ & $4$ & $1$ \\
{\bf $25\%$}   & $161$ & $46$ & $14$ & $4$ & $2$ \\
{\bf $50\%$}   & $158$ & $48$ & $14$ & $4$ & $2$ \\
{\bf $75\%$}   & $155$ & $54$ & $17$ & $5$ & $2$ \\
{\bf $100\%$} & $146$ & $56$ & $17$ & $6$ & $2$ \\
\hline\hline
\end{tabular}
\begin{tabular}{c c c c c c c}
\hline\hline
   {\it Neigh. Mode - $\%$ of rnd prom.} & {\it  $N_5$} & {\it  $N_4$} & {\it  $N_3$} & {\it $N_2$} & {\it $N_1$} \\
\hline
{\bf $0\%$}     & $133$ & $61$ & $21$ & $8$ & $2$ \\
{\bf $25\%$}   & $134$ & $63$ & $22$ & $7$ & $2$ \\
{\bf $50\%$}   & $132$ & $63$ & $23$ & $7$ & $2$ \\
{\bf $75\%$}   & $128$ & $64$ & $25$ & $6$ & $3$ \\
{\bf $100\%$} & $128$ & $65$ & $24$ & $7$ & $3$ \\
\hline\hline
\end{tabular}
\caption{{\it  With reference to the simulations plotted in Fig.12 we report, from left to right, for both 'global' (upper panel)and 'neighbors' (lower panel) modes and for all the percentages of random promotions, the total number of agents which terminated their career at each one of the $5$ levels of the organization.}}
\end{center}
\label{summary-table2}
\end{table}           
Such a result sounds paradoxical because, at first glance, the total number of promotions, $N_p$, should be strictly correlated with the total number of dismissals and retirements at levels $4,3,2,1$ (quantified by $N_4$, $N_3$, $N_2$, $N_1$). Actually, from Tables 1 and 2, it results that, for each percentage of random promotions and for both the topologies, it is always 
\begin{equation}  
N_p=N_4+(N_3\cdot2)+(N_2\cdot3)+(N_1\cdot4)
\label{eq1}
\end{equation}  
since every dismissal or retirement causes, in turn, a cascade of promotions depending on the level at which it occurs (the more numerous the higher the level). On the other hand, it is also evident that, independently of the topology, it must always be 
\begin{equation}  
N_d+N_r=N_5+N_4+N_3+N_2+N_1. 
\label{eq2}
\end{equation}  
Therefore, comparing equations (3) and (4), the explanation of why $N_p$ is so sensitive to the topology, while $N_d$ and $N_r$ are not, seems to lie in the different role played by the bottom level $5$, since only in this level a dismissal or a retirement do not translate into a promotion but imply a new external engagement. Actually, the value of $N_5$ in the 'neighbors' mode is always lower of the correspondent value in the 'global' mode (see Table 2) exactly of those about $30$ units that, as previously observed, in a modular organization are more likely to be promoted to the next levels with respect to a pyramidal structure.
\\
The solution of this puzzle becomes clear if one looks to the ages of agents at level $5$, which (during the dynamics) are normally distributed around $40$ years independently of the topology, and those at the other levels. As a matter of fact, it results that in a modular organization ('neighbors' mode) members of levels $\le 4$ are on average quite older than those of the bottom level $5$, while the same does not happen in a pyramidal topology ('global' mode). Such a strange statistical effect can be explained with an argument similar to that used in subsection $3.1$, since in 'neighbors' mode the choice of the candidate to be promoted is carried out on a much smaller sample (of the order of a few units) than in the 'global' mode, so it is very likely that the tails of the age distribution will almost always excluded, and the age of the promoted members will be closer to the average, i.e.  $40$ years old. This implies a greater number of retirements at levels $\le 4$ with respect to level $5$ in 'neighbors' mode (despite the fact that the total number $N_r$ is similar for both the topologies) and, in turn, implies the observed greater number of promotions $N_p$ for the modular topology. 
\\
In conclusion, it seems that the mere fact of adopting a modular structure, instead of a pyramidal one, increases not only the efficiency of the organization in a meritocratic regime of promotions (see the first column of Table 1), but also the probability of career advancements for its members. Therefore it seems that  a complex hierarchical and modular  topology is convenient  both from a global and an individual point of view. Finally, we notice that the total number of promotions $N_p$  grows even further  also  by increasing the percentage of random promotions for both the topologies, thus definitively confirming the advantage of using a random strategy for any kind of organization (if the Peter hypothesis holds).

\subsection{Real applications of random promotion strategies}

Let us now discuss the possibility of real applications of a random promotion strategy.
A frequent objection to the adoption of a random promotion strategy, by a company or a public administration, concerns the possible negative psychological feedback of employees to a denied and expected promotion. It is true that, for the  sake of simplicity, we did not consider any possibility of this kind in our simulations. But, on one hand, in a very big company it is very likely that the employees do completely ignore the promotion strategies of their managers, therefore these effects could have a very minimal influence on their work. On the other hand, we think that the most important point which emerges from our work is the necessity to distinguish promotions from rewards and incentives for the good work done. Actually, if the best employees understandably expect a prize for their work, such a prize, which from an individual perspective is necessary for the purpose of preventing a decrease in competence (see also the Prince Charles syndrome), does not necessarily have to coincide with a promotion. On the contrary, receiving an increase in the salary or more responsibility or more  freedom in the time schedule, could be very often a much more appreciated reward for an excellent performance. 
Thus we explicitly suggest to accompany random promotions with prize/rewards to the best members, in order to prevent psychological side effects.
As a matter of fact, it is not difficult to understand  that if someone is the best for a certain role, it is much better to try to keep him/her in that position instead of risking to change his/her task or role, a change that, if the Peter hypothesis holds, could expose the company to the risk of a decrease in efficiency. For example, it is surely much  better to keep an excellent surgeon in his/her position instead of promoting him/her as the main director of the hospital, which is a managerial role: in so doing we risk to have a double loss, i.e. a less competent new director and a less competent new surgeon!
\\
But, once established that for a given organization the meritocratic strategy is a loosing one if the Peter hypothesis holds, why random promotions should be better? Of course, an immediate answer is that promoting people at random prevents the change of role of the best members with certainty, thus circumventing the Peter effect. On the other hand, it is not easy to understand analytically this result in detail, since the effect is a cooperative emergent feature of the numerical simulations. We suspect that such an effect could have many similarities with the Parrondo paradox \cite{Parrondo}, where the noise has a constructive role. Also in physics there  are many examples where noise has a positive influence \cite{Mantegna,Frenkel,Page,Caruso1,Caruso2} and actually this was the hint that stimulated us in trying this possibility to bypass the Peter principle. 
However, apart from physics, there are other analogies in nature in favor of this strategy and one of them is for sure natural selection. In fact during the evolution, natural selection proceeds through random mutations and not through something like top-down meritocratic promotions. If a random mutation reveals to give a great advantage for some species, then it is maintained and reinforced, never changed or removed on purpose.  Moreover  mixing genes is surely  positive  for   species health and survival. 
\\
In addition to these arguments, a random strategy of promotions has also other advantages which provide further benefits. In fact a random selection can favor the emergence of hidden skills of the less competent employees, which otherwise could have had very few probabilities to be appreciated. A famous  example in this direction, extracted from another field, is the case of the well-known Opera singer Maria Callas. Her great turning point in career occurred in $1949$ in Venice when she was chosen, by chance and all of a sudden, to substitute the main singer, who had fallen ill \cite{Callas}, in a role she had never interpreted, in the opera "I Puritani". 
Something similar happened also to the famous orchestra conductor Arturo Toscanini, who debuted as a conductor at the age of $19$ years old just by chance, forced to substitute the official one who abandoned the orchestra (where Toscanini played as musician) during a tourn\'ee in south America\cite{Toscanini}.  
\\
But  there are also other  justifications for the practical use of a random selection of candidates. Not infrequently, especially in the public administration, the meritocratic regime does not represent the main criterium chosen for promoting people: relatives or friends with very minimal competence are often preferred for promotion to higher levels, without any relation to competence. In this case a random selection of candidates may have the merit of disrupting this very bad practice which for sure decreases the organization efficiency more than any meritocratic promotion. In this respect,  Phedon Nicolaides \cite{Nicolaides}, after reading our first paper, recently expressed his personal point view on our proposal by considering a random choice for the committees that should select employees for promotion. In our opinion this could also be a successful strategy, although we have not done any simulation in this direction. 
\\   
At this point it would be interesting to put in practice and test our proposal in the real world to have an experimental proof, beyond simulations, of its validity and  we hope  to collaborate with private companies in this respect in the near future. Of course, in order to do so, our proposal needs to be adapted to real organizations, with a given topology, which could differ from case to case. But we have recently discovered that, as far as we know, there is at least a well known case where a strategy similar to what we propose has been put in practice with success. This is the case of the SEMCO company in Brazil \cite{Semco}. In the 80's this company was saved from bankruptcy by Ricardo Semler, who took the role of CEO substituting his father. Semler, by adopting a very innovative way of management based on tasks rotation, on a system of rewards and on a very limited role of hierarchy, invented a new strategy which revealed to be very successful and increased the number of employees from 90 to 3000. Today Semler is one of the new guru of management who gives lectures at the  Harvard Business School and in other prestigious institutions around the world. The new Semco way is not exactly equivalent to what we propose, but it is very similar in many respects and provides an interesting  example of how new ways of management, apparently eccentric and not based on a simple and naively meritocratic regime, are not only possible but, above all, convenient. 
\\
Finally, before closing this discussion, we would like to mention another possible application of our random strategy to political elections. When democracy was invented in the old times in Athens \cite{democrazia}
political representatives were sorted at random and not elected. An election  system based on a refined lottery was adopted in the past for choosing  the Doge in Venice  \cite{doge} and similar systems were adopted in the past in other parts of Italy. So it would be quite  interesting to see if a random selection of political representative would really improve the efficiency of a parliament. This idea has regained some strength also recently with the proposal in this direction of popular juries, that should control the work of politicians, by S\'egol\'ene Royal in France \cite{Royal}, or with the proposal of Barnett and Carty for a radical reform of the House of Lords by a random election \cite{Barnett}. Stimulated by these examples we have started to work in this direction with numerical simulations. Our results seem very encouraging and favor the adoption of a random choice of representatives for enhancing the efficiency of a parliament, but they will be reported elsewere\cite{parlamento}.

\section{Conclusions}
We have explored in a more realistic model of hierarchical organization, within an agent-based simulation approach, the negative effects of the Peter principle and possible strategies to contrast them.  Our results confirm the robustness of our previous findings, providing further support to our claim that a random strategy seems to be a very simple and interesting way to increase the efficiency of any real organization for which the Peter hypothesis of uncorrelated competence transmission holds. As discussed in the paper, the random selection not necessarily should be completely random to be successful, since a percentage of even $50\%$ of random selection provides already a good advantage with respect to a full "naively" meritocratic regime of promotions.  
On the other hand, it is also true that no psychological effects of random promotions have been taken into account in our simulations and that these could be non-negligible for real organizations. We will surely consider them in future studies, but, anyway, we believe that a random selection with different possible refinements (as suggested for example in \cite{Nicolaides}) could be  an interesting  promotion strategy to improve the performance  of a hierarchical organization, not only to contrast  the Peter principle effects  but also nepotism and corruption. In general a rotation of tasks, which is a very similar possibility to the one we propose, has already been applied successfully    at least in a real case, i.e. the SEMCO company \cite{Semco}.   
We therefore believe that our study is not merely an academic out-of-the world proposal and  could have  useful and relevant practical implications. In this respect, an experimental test in a real system is very welcome and we hope to present soon, in a future study, some first concrete results in this direction.

\section{Acknowledgements}
We would like to thank P. Sobkowicz for useful discussions and M. Abrahams for calling our attention to  
Ref.\cite{Phelan-Lin}.

\end{document}